\documentclass{aa}
\usepackage{txfonts}
\usepackage{graphicx}
\usepackage{subcaption}
\usepackage[colorlinks=true,allcolors=blue]{hyperref}
\usepackage{natbib}

\begin{document} 

\titlerunning{}
\authorrunning{Mountrichas \& Georgantopoulos}
\titlerunning{The properties of SMBH and their host galaxies for type 1 and 2 AGN}

\title{The properties of supermassive black holes and their host galaxies for type 1 and 2 AGN in the eFEDS and COSMOS fields}

\author{G. Mountrichas\inst{1} and I. Georgantopoulos\inst{2}}
          
    \institute {Instituto de Fisica de Cantabria (CSIC-Universidad de Cantabria), Avenida de los Castros, 39005 Santander, Spain
              \email{gmountrichas@gmail.com}
           \and
             National Observatory of Athens, Institute for Astronomy, Astrophysics, Space Applications and Remote Sensing, Ioannou Metaxa
and Vasileos Pavlou GR-15236, Athens, Greece
}

\abstract{In this study, our primary objective is to compare the properties of supermassive black holes (SMBH) and their host galaxies between type 1 and type 2 active galactic nuclei (AGN). In our analysis, we use X-ray detected sources in two fields, namely the eFEDS and the {\it{COSMOS-Legacy}}. To classify the X-ray sources, we perform spectral energy distribution (SED) fitting analysis, using the CIGALE code. Ensuring the robustness of our analysis is paramount, and to achieve this, we impose stringent selection criteria. Thus, only sources with extensive photometric data across the optical, near- and mid-infrared part of the spectrum and reliable host galaxy properties and classifications were included. The final sample consists of 3,312 AGN, of which 3\,049 are classified as type 1 and 263 as type 2. The sources span a redshift range of $\rm 0.5<z<3.5$ and encompass a wide range of X-ray luminosities, falling within $\rm 42<log,[L_{X,2-10keV}(ergs^{-1})]<46$. Our results show that type 2 AGN exhibit a tendency to inhabit more massive galaxies, by $0.2-0.3$\,dex (in logarithmic scale), compared to type 1 sources. Type 2 AGN also display, on average, lower specific black hole accretion rates, a proxy of the Eddington ratio, compared to type 1 AGN. These differences persist across all redshifts and L$_X$ considered within our dataset. Moreover, our analysis uncovers, that type 2 sources tend to have lower star-formation rates compared to 1 AGN, at $\rm z<1$. This picture reverses at $\rm z>2$ and $\rm log,[L_{X,2-10keV}(ergs^{-1})]>44$. Similar patterns emerge when we categorize AGN based on their X-ray obscuration levels ($N_H$). However, in this case, the observed differences are pronounced only for low-to-intermediate L$_X$ AGN and are also sensitive to the $\rm N_H$ threshold applied for the AGN classification. These comprehensive findings enhance our understanding of the intricate relationships governing AGN types and their host galaxy properties across diverse cosmic epochs and luminosity regimes.}

\keywords{}
   
\maketitle

\section{Introduction}

Active galactic nuclei (AGN) occupy a pivotal role in the process of galaxy evolution. These AGN derive their energy from the accretion of matter onto the supermassive black hole (SMBH) situated at the heart of galaxies. The co-evolution of AGN and their host galaxies is intricately regulated by mechanisms governing SMBH accretion and the subsequent feedback from AGN. In order to unravel this complex interaction between the active SMBH and its host galaxy, it becomes imperative to gain insights into the internal structure of AGN. A fundamental aspect of this endeavor involves elucidating the physical distinctions that underlie obscured and unobscured AGN.

According to the unification model \citep[e.g.,][]{Urry1995, Nenkova2002, Netzer2015} an AGN is classified as obscured or unobscured based on the angle of our line of sight relative to the symmetry axis of the accretion disk and torus surrounding the central black hole. When we observe the AGN edge-on, we classify it as obscured, whereas an AGN is unobscured when is viewed face-on. As the understanding of AGN structures has evolved, more intricate models \citep{Ogawa2021, EsparzaArredondo2021} have emerged to accommodate diverse classifications observed across various wavelengths. These models aim to account for the variety of classifications across different wavelengths \citep[for instance, distinctions between X-ray and optical classifications; ][]{OrdovasPascual2017}. However, despite these advancements, the primary determinant for discerning obscured and unobscured AGN, according to the unification model, continues to be the inclination angle.

An alternative interpretation of AGN obscuration arises within the realm of evolutionary models. According to this perspective, the distinct AGN types are attributed to the fact that SMBHs and their host galaxies are observed during different evolutionary phases. The core concept underlying these models posits that obscured AGN are observed during an early phase when the energy output generated by the accretion disk surrounding the SMBH is not sufficiently robust to disperse the surrounding gas. As material continues to accrete onto the SMBH, its energy output intensifies, eventually compelling the obscuring material to dissipate \citep[e.g.,][]{Ciotti1997, Hopkins2006}.

Obtaining a deeper understanding of the characteristics of both AGN populations is crucial to unveil various facets of the intricate relationship between AGNs and galaxies. One commonly adopted method to accomplish this is to compare the host galaxy properties of obscured and unobscured AGN. If the two populations live in similar environments, this would provide support to the unification model whereas if they reside in galaxies of different properties, it would suggest that they are observed at different evolutionary phases.

AGN can be classified into obscured and unobscured through diverse criteria. For example, the classification into these two groups can be accomplished using X-ray criteria, such as relying on the hydrogen column density or the hardness ratio \citep{Mountrichas2020}. Furthermore, optical spectral features contribute to the classification of AGN into type 1 (unobscured) and type 2 (obscured). Type 1 AGN exhibit broad lines in their optical spectra, while type 2 AGN lack these broad emission lines \citep{Zou2019}. Notably, intermediate optical spectroscopic classifications, such as sub-types 1.0, 1.2, 1.5, 1.8, and 1.9, are also viable within this framework \citep{Whittle1992}. Another classification method involves categorizing AGN into type 1 and type 2 based on their inclination angle, $i$, determined through spectral energy distribution (SED) fitting \citep{Mountrichas2022b}. In this case, the sub-categorization of AGN is not possible since the calculation of $i$ through SED fitting analysis, is not sensitive to incremental changes of $i$ \citep[e.g.,][]{Yang2020}. It is well-acknowledged that the classification of AGN may vary when employing these distinct criteria \citep[e.g.][]{Masoura2020, Mountrichas2022b}.

Prior investigations that relied on optical criteria, such as optical spectra, to classify X-ray AGN into type 1 and 2 found that type 2 sources tend to inhabit more massive systems compared to type 1, but no statistically significant differences were found regarding the star-formation rate (SFR) of the two AGN populations. It is important to note, though, that these earlier studies either concentrated on AGN primarily within the realm of low-to-moderate X-ray luminosities, $\rm log\,[L_{X,2-10keV}(ergs^{-1})]<44$, AGN \citep{Zou2019} and/or they restricted their analysis to low redshifts \citep[$\rm z<1$;][]{Mountrichas2021b}.

Alternatively, AGN are classified into obscured and unobscured using X-ray criteria. For instance, using the hydrogen column density , $\rm N_H$, and a threshold at $\rm N_H=10^{21.5}$\,cm$^{-2}$ (or $\rm N_H=10^{22}$\,cm$^{-2}$), previous studies found no significant differences regarding the SFR and stellar mass, M$_*$, of galaxies hosting X-ray absorbed and unabsorbed AGN \citep[e.g.,][]{Masoura2021, Mountrichas2021c}. More recently, \cite{Georgantopoulos2023}, adopted a higher $\rm N_H$ threshold ($\rm N_H=10^{23}$\,cm$^{-2}$) and found that the two AGN populations live in galaxies with statistically significant differences ($>2\sigma$) in terms of their SFR and M$_*$. Moreover, X-ray absorbed sources tend to exhibit lower specific black hole accretion rates, $\lambda_{sBHAR}$, which serves as a proxy for the Eddington ratio, compared to their unabsorbed counterparts. They attributed these divergent findings compared to previous investigations to either the elevated $\rm N_H$ threshold employed or the varying X-ray luminosities probed by the datasets used across different studies. Mountrichas et al. (submitted) built upon these insights by employing X-ray AGN data from the 4XMM catalog and implementing rigorous X-ray criteria for AGN classification. Their results closely mirrored those obtained by \cite{Georgantopoulos2023}, reinforcing the observed differences in host galaxy properties between X-ray obscured and unobscured AGN. Moreover, their analysis highlighted that these distinctions tend to diminish for luminous AGN. However, it's worth noting that the application of X-ray and optical criteria for AGN categorization may not always yield identical classification \citep[e.g.,][]{Masoura2020, Mountrichas2020}.

In this work, we use X-ray AGN detected in eROSITA Final Equatorial Depth Survey (eFEDS)
and the {\it{COSMOS-Legacy}} fields. Our primary objective revolves around comparing the SMBH and host galaxy properties of type 1 and 2 AGN. Sources are classified using the outcomes from applying SED fitting analysis, employing the CIGALE code. Specifically, the measurement of the inclination angle, provided by CIGALE is used for the AGN categorization. Sect. 3 presents the SED fitting analysis, the photometric data and the quality selection criteria applied. These criteria were designed to identify sources with robust host galaxy measurements and reliable classifications. Additionally, the section elaborates on the criteria utilized to differentiate AGN into the two distinct types. The presentation and discussion of our findings  are provided in Sections 4 and 5, respectively. We summarize our main conclusions in Sect. 6.


\section{Data}
\label{sec_data}

In our analysis, we use X-ray AGN detected in the eFEDS and COSMOS fields. Both datasets and the quality  criteria applied are described in detail in sections 2 of \cite{Mountrichas2022b}  and \cite{Mountrichas2022a}. Below we present a brief summary.

\subsection{X-ray sources in the eFEDS field}
\label{sec_data_efeds}

The eFEDS X-ray catalogue includes 27910 X-ray sources detected in the $0.2-2.3$\,keV energy band with detection likelihoods $\geq 6$, that corresponds to a flux limit of $\approx 7 \times 10^{-15}$\,erg\,cm$^{-2}\,\rm s{^{-1}}$ in the $0.5-2.0$\,keV energy range \citep{Brunner2022}. \cite{Salvato2022} presented the multiwavelength counterparts and redshifts of the X-ray sources, by identifying their optical counterparts. Two independent methods were utilized to find the counterparts of the X-ray sources, {\tiny {NWAY}} \citep{Salvato2018b} and {\tiny {ASTROMATCH}} \citep{Ruiz2018}. {\tiny {NWAY}} is based on Bayesian statistics and {\tiny {ASTROMATCH}} on the Maximum Likelihood Ratio \citep{Sutherland_and_Saunders1992}. For $88.4\%$ of the eFEDS point like sources, the two methods point at the same counterpart. Each counterpart is assigned a quality flag, {\textsc{CTP\_quality}}. Counterparts with $\rm {\textsc{CTP\_quality}} \geq 2$ are considered reliable, in the sense that either both methods agree on the counterpart and have assigned a counterpart probability above threshold ($\rm {\textsc{CTP\_quality}} = 4$ for 20873 sources), or both methods agree on the counterpart but one method has assigned a probability above threshold ($\rm {\textsc{{\textsc{CTP\_quality}}}} = 3$, 1379 sources), or there is more than one possible counterparts ($\rm {\textsc{{\textsc{CTP\_quality}}}} = 2$, 2522 sources). Only sources with $\rm {\textsc{{\textsc{CTP\_quality}}}} \geq 3$ are included in our analysis. Moreover, \cite{Mountrichas2022b} cross-matched the X-ray dataset with the GAMA-09 photometric catalogue produced by the HELP collaboration \citep{Shirley2019, Shirley2021}, that covers $\sim 35\%$ of the eFEDS area to extend the photometric coverage to far-infrared wavelengths. HELP provides data from 23 extragalactic survey fields, imaged by the {\it{Herschel}} Space Observatory which form the {\it{Herschel}} Extragalactic Legacy Project (HELP). About $\sim 10\%$ of the X-ray sources in the eFEDS field have available {\it{Herschel}}/SPIRE photometry. 

eFEDS has been observed by a number of spectroscopic surveys, such as GAMA \citep{Baldry2018}, SDSS \citep{Blanton2017} and WiggleZ \citep{Drinkwater2018}. Only sources with secure spectroscopic redshift, {\textit{specz}}, from the parent catalogues were considered in the eFEDS catalogue \citep{Salvato2022}. 6640 sources have reliable {\textit{specz}}. Photometric redshifts, {\textit{photoz}}, were computed for the remaining sources using the LePHARE code \citep{Arnouts1999, Ilbert2006} and following the procedure outlined in e.g., \cite{Salvato2009, Salvato2011}. A redshift flag is assigned to each source, CTP\_REDSHIFT\_GRADE. Only sources with $\rm CTP\_REDSHIFT\_GRADE \geq 3$ (26047/27910) are considered in this work. This criterion includes sources with either spectroscopic redshift ($\rm CTP\_REDSHIFT\_GRADE = 5$) or the {\textit{photoz}} estimates of the two methods agree ($\rm CTP\_REDSHIFT\_GRADE = 4 $) or agree within a tolerance level \citep[$\rm CTP\_REDSHIFT\_GRADE = 3$; for more details see Sect. 6.3 of][]{Salvato2022}.

\cite{Liu2022} performed a systematic X-ray spectral fitting analysis on all the X-ray systems. Based on their results only $10\%$ of the sources are X-ray obscured. In this work, we use their posterior median, intrinsic (absorption corrected) X-ray fluxes in the $2-10$\,keV energy band.

\subsection{X-ray sources in the COSMOS field}
\label{sec_data_cosmos} 

To increase the size of the sample used in our analysis, and in particular the number of type 2 AGN, we also add sources detected in the {\it{COSMOS-Legacy}} survey \citep{Civano2016}. {\it{COSMOS-Legacy}}  is a 4.6\,Ms {\it{Chandra}} program that covers 2.2\,deg$^2$ of the COSMOS field \citep{Scoville2007}. The central area has been observed with an exposure time of $\approx 160$\,ks while the remaining area has an exposure time of $\approx 80$\,ks. The limiting depths are $2.2 \times 10^{-16}$, $1.5 \times 10^{-15}$ , and $8.9 \times 10^{-16}\,\rm erg\,cm^{-2}\,s^{-1}$ in the soft (0.5-2\,keV), hard (2-10\,keV), and full (0.5-10\,keV) bands, respectively. The X-ray catalogue includes 4016 sources. We only use sources within both the COSMOS and UltraVISTA \citep{McCracken2012}  regions. UltraVISTA covers 1.38\,deg$^2$ of the COSMOS field \citep[][]{Laigle2016} and has deep near-infrared (NIR) observations ($J, H, K_s$ photometric bands) that allow us to derive more accurate host galaxy properties through SED fitting (see below). There are 1718 X-ray sources that lie within the UltraVISTA area of COSMOS.

\cite{Marchesi2016} matched the X-ray sources with optical and infrared counterparts using the likelihood ratio technique \citep{Sutherland_and_Saunders1992}. Of the sources, 97\%  have an optical and IR counterpart and a {\textit{photoz}} and $\approx 54\%$  have {\textit{specz}}. {\textit{photoz}} available in their catalogue, have been produced following the procedure described in \cite{Salvato2011}. The accuracy of photometric redshifts is found at $\sigma_{\Delta z/(1+z_{spec})}=0.03$. The fraction of outliers ($\Delta z/(1+z_{zspec})>0.15$) is $\approx 8\%$. \cite{Mountrichas2022a} cross-matched the X-ray catalogue with the COSMOS photometric dataset produced by the HELP collaboration to assign far-infrared photometry to the sources. About $\sim 60\%$ of the X-ray sources in the UltraVISTA region have available {\it{Herschel}}/SPIRE photometry.

The catalogue presented in \cite{Marchesi2016} also provides measurements of the intrinsic column density, N$\rm _H$, estimated using hardness ratios ($\rm HR=\frac{H-S}{H+S}$, where H and S are the net counts of the sources in the hard and soft band, respectively) and the method (Bayesian estimation of hardness ratios, BEHR) presented in \cite{Park2006}. An X-ray spectral power law with slope $\Gamma=1.8$ is also assumed.

\section{Analysis}
\label{sec_analysis}

In this section, we outline the methodology employed to measure the host galaxy properties of the X-ray sources and describe the criteria utilized for the selection of sources with the most robust measurements and reliable classification.

\subsection{Host galaxy properties}

The host galaxy properties of the X-ray AGN have been calculated via SED fitting, using the CIGALE code \citep{Boquien2019, Yang2020, Yang2022}. The SED fitting analysis is described in detail in sections 3.1 in \cite{Mountrichas2022a} and \cite{Mountrichas2022b}, for sources in the COSMOS and eFEDS fields, respectively. 

In brief, the galaxy component is modelled using a delayed SFH model with a function form $\rm SFR\propto t \times exp(-t/\tau)$. A star formation burst is included \citep{Malek2018, Buat2019} as a constant ongoing period of star formation of 50\,Myr. Stellar emission is modelled using the single stellar population templates of \cite{Bruzual_Charlot2003} and is attenuated following the \cite{Charlot_Fall_2000} attenuation law. To model the nebular emission, CIGALE adopts the nebular templates based on \cite{VillaVelez2021}. The emission of the dust heated by stars is modelled based on \cite{Dale2014}, without any AGN contribution. The AGN emission is included using the SKIRTOR models of \cite{Stalevski2012, Stalevski2016}. The parameter space used in the SED fitting process is shown in Tables 1 in \cite{Mountrichas2022a, Mountrichas2022b}. CIGALE has the ability to model the X-ray emission of galaxies. In the SED fitting process, the intrinsic L$_X$ in the $2-10$\,keV band, provided in the \cite{Marchesi2016}, for the COSMOS dataset, and \cite{Liu2022}, for the eFEDS sample, are used. The reliability of the SFR measurements  has been examined in detail in our previous works and, in particular, in Sect. 3.2.2 in \cite{Mountrichas2022a}.

\subsection{Selection of AGN with robust SED fitting measurements} 

In order to get reliable SED fitting results, it is essential to restrict the analysis to those sources with the highest possible photometric coverage. For that purpose, \cite{Mountrichas2022a} and \cite{Mountrichas2022b} required the X-ray AGN to have available the following photometric bands $u, g, r, i, z, J, H, K$, W1/IRAC1, W2/IRAC2, W4/MIPS24, where W1, W2, W4 are the photometric bands of WISE \citep{Wright2010}, at 3.4\,$\mu m$, 4.6\,$\mu m$ and 22]\,$\mu m$, respectively, and IRAC1, IRAC2 and MIPS24 are the 3.6\,$\mu m$, 4.5\,$\mu m$ and 24\,$\mu m$ photometric bands of {\it{Spitzer}}. They also applied the following requirements in the CIGALE's results: a reduced $\chi ^2$  threshold of $\chi ^2_r <5$ was imposed \citep[e.g.][]{Masoura2018, Buat2021} and sources for which CIGALE could not constrain the parameters of interest (SFR, M$_*$) were excluded from the analysis. Specifically, CIGALE provides two values for each estimated galaxy property. One value corresponds to the best model and the other value (bayes) is the likelihood-weighted mean value. A large difference between the two calculations suggests a complex likelihood distribution and important uncertainties. Therefore, in our analysis were included only sources for which $\rm \frac{1}{5}\leq \frac{SFR_{best}}{SFR_{bayes}} \leq 5$ and $\rm \frac{1}{5}\leq \frac{M_{*, best}}{M_{*, bayes}} \leq 5$, where SFR$\rm _{best}$ and  M$\rm _{*, best}$ are the best-fit values of SFR and M$_*$, respectively and SFR$\rm _{bayes}$ and M$\rm _{*, bayes}$ are the Bayesian values estimated by CIGALE.

In the SED fitting analysis followed by \cite{Mountrichas2022b}, they have used the Gaussian Aperture and Photometry (GA{\tiny {A}}P) photometry that is available in the eFEDS X-ray catalogue. GA{\tiny {A}}P photometry has been performed twice, with aperture setting $\rm MIN\_APER= 0\arcsec.7$ and 1\arcsec.0. A value for each photometric band with the optimal MIN\_APER is provided \citep[for the choice of GA{\tiny {A}}P aperture size, see ][]{Kuijken2015}. GA{\tiny {A}}P is optimised for calculating {\textit{photoz}} that require colour measurements. In the case of extended and low redshift sources, total fluxes may be underestimated \citep{Kuijken2019}. Due to these considerations, they have opted to omit sources with low redshift values (z < 0.5) from their analysis. We adhere to their rationale and also implement the identical redshift limit in our analysis. For consistency, the same requirement is applied on the X-ray AGN in the COSMOS field.

Application of the criteria above and those mentioned in Sect. \ref{sec_data} results in 7\,279 AGN. Out of them 2\,727 ($37\%$) have spectroscopic redshift. From the 7\,279, 6131 ($84\%$) are detected in the eFEDs field and 1\,148 in the UltraVISTA region of COSMOS.

\subsection{Classification of AGN}
\label{sect_classif}

\cite{Mountrichas2021b} used X-ray AGN in the XMM-XXL field \citep{Pierre2016} and showed that CIGALE can reliably classify sources into type 1 and 2. Specifically, they identified type 1 and 2 AGN, using the bayes and best estimates of the $i$ parameter, derived by CIGALE. Type 1 AGN, based on the SED analysis, are those with $i_{best}=30^{\circ}$ and $i_{bayes}<40^{\circ}$, while secure type 2 sources are those with $i_{best}=70^{\circ}$ and $i_{bayes}>60^{\circ}$. Then, they compared CIGALE's classification with that provided in the catalogue presented in \cite{Menzel2016}, in which AGN have been divided into broad (type 1) and narrow (type 2) line sources, using the full width half maximum (FWHM) for emission lines originating from different regions of the AGN (H$_\beta$, MgII, CIII and CIV).

The analysis presented in \cite{Mountrichas2021b} revealed that the SED fitting algorithm classified type 1 AGN with an accuracy of $\sim 85\%$. A similar percentage was found regarding the completeness at which type 1 sources were identified. For type 2 sources, the performance of CIGALE was at $\sim 50\%$, both regarding the reliability and the completeness. We note that the reliability is defined as the fraction of the number of type 1 (or type 2) sources classified by the SED fitting that are similarly classified by optical spectra. The completeness refers to how many sources classified as type 1 (or type 2) based on optical spectroscopy were identified as such by the SED fitting results. Therefore, for the purposes of this work, we are, mainly, interested in the reliability performance of CIGALE. 

The reliability of $\sim 85\%$ with which CIGALE identifies type 1 sources is acceptable for the purposes of our statistical analysis. However, the reliability of the SED fitting code regarding type 2 AGN is rather low, since it implies that about half of the sources identified as type 2 by CIGALE are, in fact, misclassified sources. Nevertheless, \cite{Mountrichas2021b} showed that the vast majority ($\sim 82\%$) of the misclassified type 2 sources have increased polar dust values (E$_{B-V}>0.15$; see their Fig. 8 and Sect. 5.1.1). Thus, we exclude these sources from our analysis and classify as type 2 those AGN that meet the inclination angle criteria mentioned above and also have polar dust values lower than E$_{B-V}<0.15$. We note, that the introduction of polar dust in the fitting process improves the accuracy of CIGALE in the source type classification, in particular, in its reliability to identify type 2 sources \citep[see Sect. 5.5 in][]{Mountrichas2021a}. 

Moreover, it is essential to emphasize that not all of the excluded type 2 sources would be categorized as type 1 based on optical spectra. For example, in the study by \cite{Mountrichas2021b}, a substantial proportion of sources that CIGALE identified as type 2 and exhibited elevated polar dust content were subsequently confirmed as type 2 through spectroscopic classification ($\frac{9}{32}=28\%$). However, as previously mentioned, the exclusion of these systems enhances the reliability of CIGALE in its ability to discern type 2 AGN, as supported by the findings in \cite{Mountrichas2021b}.

\begin{figure}
\centering
  \includegraphics[width=0.95\columnwidth, height=7.2cm]{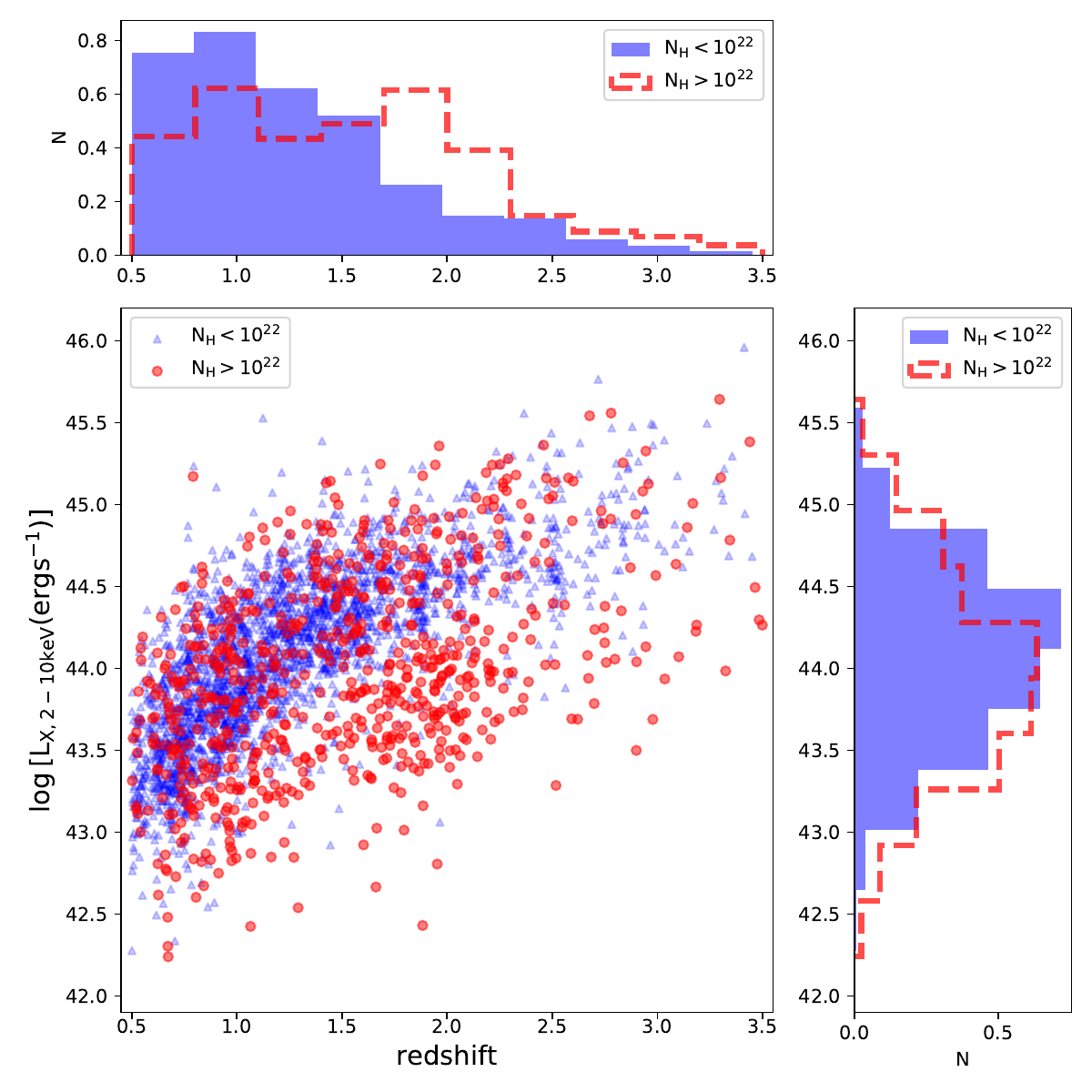}   
  \caption{Distribution of type 1 (blue triangles) and type 2 (red circles) AGN used in our analysis, in the L$_X-$redshift plane. }
  \label{fig_lx_redz}
\end{figure}

Application of these criteria on the 7\,249 AGN (see previous section) results in 3,312 reliably classified sources (Table \ref{table_classified}). 3\,049 of them are type 1 (2\,696 in eFEDS and 353 in COSMOS) and 263 are type 2 (147 in eFEDS and 116 in COSMOS). Their distribution in the L$_X-$redshift plane is shown in Fig. \ref{fig_lx_redz}. These are the sources we use in our analysis.

\begin{table}
\caption{Number of type 1 and 2 AGN, classified by CIGALE, based on the inclination angle measurements (see text for more details).}
\centering
\setlength{\tabcolsep}{3mm}
\begin{tabular}{ccc}
 \hline
{field} & {type 1} & {type 2} \\
 \hline
eFEDS  & 2\,696 & 147  \\
COSMOS  & 353 & 116  \\
  \hline
Total  &  3\,049 & 263 \\
   \hline
\label{table_classified}
\end{tabular}
\end{table}

\begin{table*}
\caption{Weighted median values of SFR, M$_*$ and $\lambda _{sBHAR}$ for type 1 and 2 sources, in the three redshift intervals used in our analysis. The number of sources and the p$-$values yielded by applying KS$-$tests, are also presented.}
\centering
\setlength{\tabcolsep}{3mm}
\begin{tabular}{ccccccc}
 \hline
 & \multicolumn{2}{c}{0.5<z<1.0} &  \multicolumn{2}{c}{1.0<z<2.0} &  \multicolumn{2}{c}{2.0<z<3.5}\\
 \hline
  & type 1 & type 2 & type 1 & type 2  & type 1 & type 2 \\
 \hline
number of sources & 1178 & 33 & 1515 & 135 & 360 & 96  \\ 
 \hline
log\,SFR  &  1.60 & 0.76 & 2.16 & 1.86 & 2.51 & 2.87 \\
p$-$value (SFR) & \multicolumn{2}{c}{0.03} & \multicolumn{2}{c}{0.08} & \multicolumn{2}{c}{0.58} \\
 \hline
log\,M$_*$ & 10.89 & 11.04 & 11.07 & 11.28 & 11.24 & 11.53 \\
p$-$value (M$_*$) & \multicolumn{2}{c}{0.17} & \multicolumn{2}{c}{0.18} & \multicolumn{2}{c}{0.17} \\
 \hline
log\,$\lambda _{sBHAR}$ & -1.29 & -1.55 & -0.86 & -1.10 & -0.55 & -0.86 \\
p$-$vaule ($\lambda _{sBHAR}$) & \multicolumn{2}{c}{0.09} & \multicolumn{2}{c}{0.86} & \multicolumn{2}{c}{0.09}\\
  \hline
\label{table_median}
\end{tabular}
\end{table*}

\begin{table*}
\caption{Same as in Table \ref{table_median}, but now splitting the X-ray dataset into two L$_X$ bins, utilizing a threshold at $\rm log\,[L_{X,2-10keV}(ergs^{-1})]=44$.}
\centering
\setlength{\tabcolsep}{2mm}
\begin{tabular}{ccccccccccccc}
 \hline
 & \multicolumn{4}{c}{0.5<z<1.0} &  \multicolumn{4}{c}{1.0<z<2.0} &  \multicolumn{4}{c}{2.0<z<3.5}\\
 \hline
 & \multicolumn{2}{c}{log\,L$_X<44$} & \multicolumn{2}{c}{log\,L$_X>44$} & \multicolumn{2}{c}{log\,L$_X<44$} & \multicolumn{2}{c}{log\,L$_X>44$} & \multicolumn{2}{c}{log\,L$_X<44$} & \multicolumn{2}{c}{log\,L$_X>44$}\\

  & type 1 & type 2 & type 1 & type 2  & type 1 & type 2 & type 1 & type 2& type 1 & type 2& type 1 & type 2\\
 \hline
number of sources & 918 & 25 & 258 & 8 & 412 & 49 & 1103 & 86 & 46 & 13 & 314 & 83\\ 
 \hline
log\,SFR  &  1.56 & 0.63 & 1.75 & 0.94 & 1.74 & 1.21 & 2.31 & 2.25 & 1.69 & 1.26 & 2.58 & 2.98\\
p$-$value (SFR) & \multicolumn{2}{c}{0.19} & \multicolumn{2}{c}{0.05} & \multicolumn{2}{c}{0.15} & \multicolumn{2}{c}{0.09} & \multicolumn{2}{c}{0.03} & \multicolumn{2}{c}{0.05} \\
 \hline
log\,M$_*$ & 10.90 & 10.95 & 10.88 & 11.04 & 11.05 & 11.28 & 11.07 & 11.29 & 11.20 & 11.31 & 11.25 & 11.58\\
p$-$value (M$_*$) & \multicolumn{2}{c}{0.42} & \multicolumn{2}{c}{0.13} & \multicolumn{2}{c}{0.52} & \multicolumn{2}{c}{0.29}  & \multicolumn{2}{c}{0.12} & \multicolumn{2}{c}{0.43} \\
 \hline
log\,$\lambda _{sBHAR}$ & -1.40 & -1.58 & -0.72 & -0.86 & -1.35 & -1.58  & -0.65 & -0.89 & -1.33 & -1.52 & -0.48 & -0.81\\
p$-$vaule ($\lambda _{sBHAR}$) & \multicolumn{2}{c}{0.62} & \multicolumn{2}{c}{0.64} & \multicolumn{2}{c}{0.29} &  \multicolumn{2}{c}{0.52} &  \multicolumn{2}{c}{0.22} &  \multicolumn{2}{c}{0.11}\\
  \hline
\label{table_median_lx}
\end{tabular}
\end{table*}

\begin{figure}
\centering
  \includegraphics[width=0.95\columnwidth, height=6.5cm]{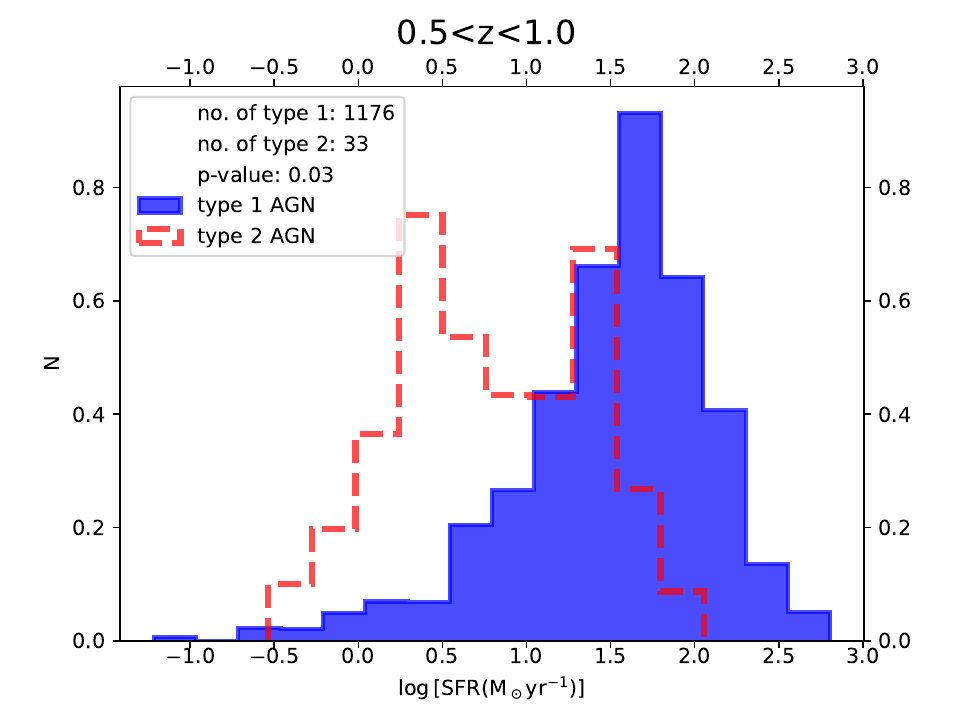}
  \includegraphics[width=0.95\columnwidth, height=6.5cm]{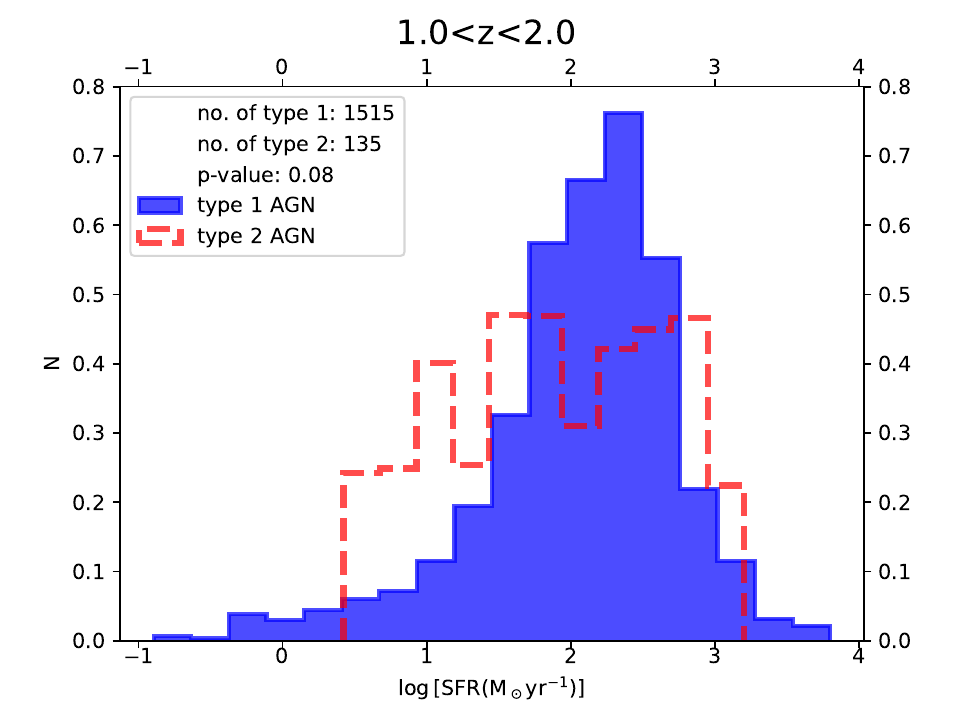}   
  \includegraphics[width=0.95\columnwidth, height=6.5cm]{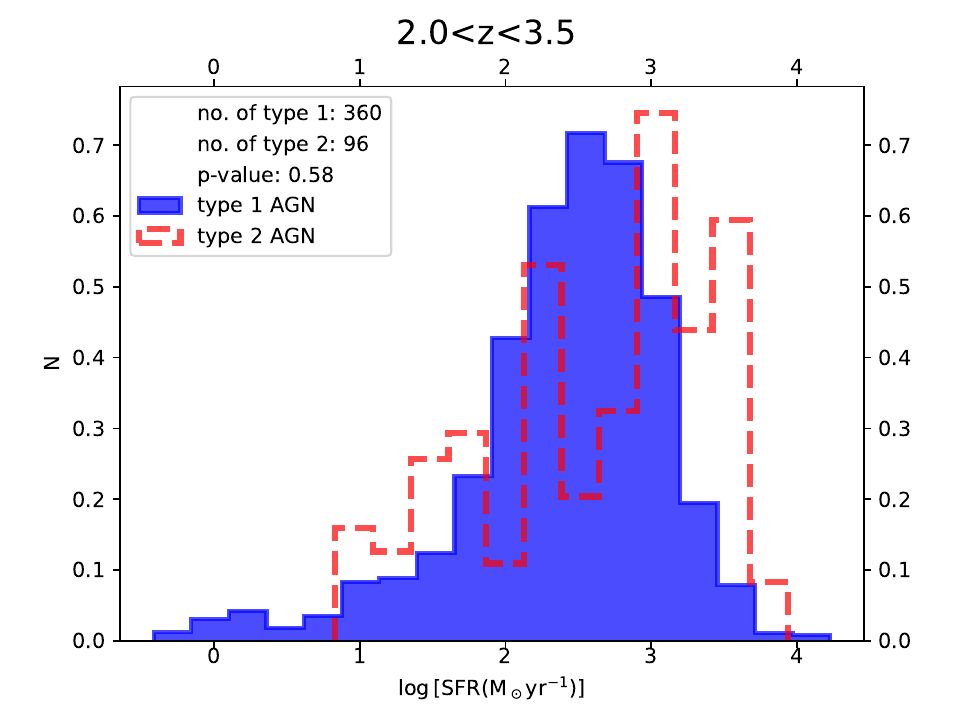}   
  \caption{Distributions of SFR of type 1 (blue shaded histogram) and type 2 (red, dashed histogram), at different redshift intervals, as indicated in the title of each panel. Distributions are weighted to account for the different L$_X$ and redshift of the two AGN populations (see text for more details).}
  \label{fig_sfr_redz}
\end{figure}

\begin{figure}
\centering
  \includegraphics[width=0.95\columnwidth, height=6.5cm]{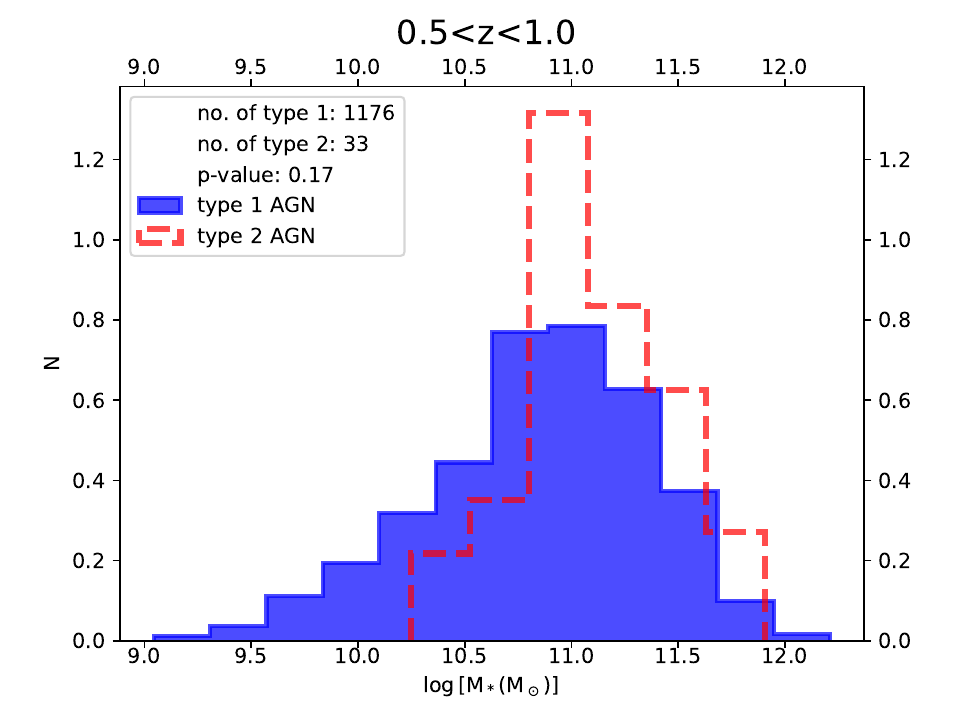} 
  \includegraphics[width=0.95\columnwidth, height=6.5cm]{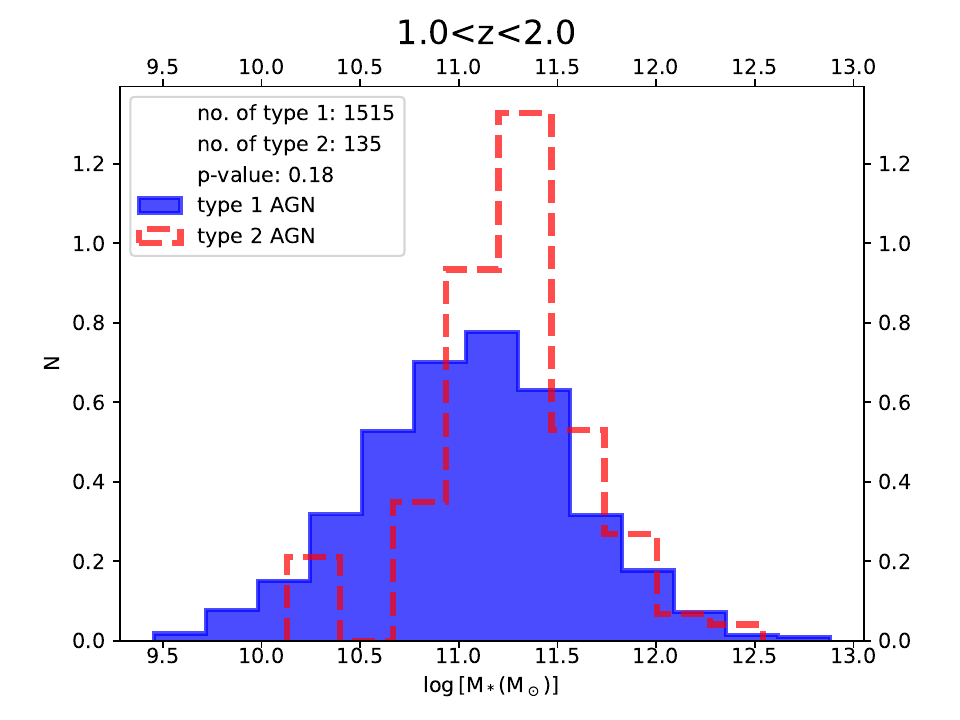} 
  \includegraphics[width=0.95\columnwidth, height=6.5cm]{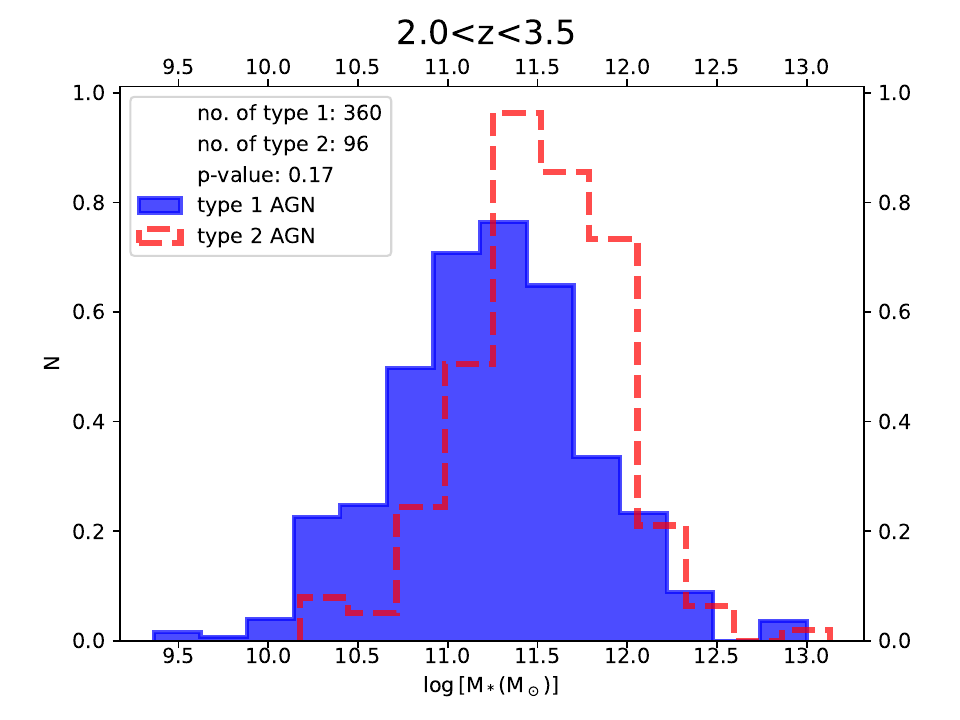} 
  \caption{Distributions of M$_*$ of type 1 (blue shaded histogram) and type 2 (red, dashed histogram), at different redshift intervals, as indicated in the title of each panel. Distributions are weighted to account for the different L$_X$ and redshift of the two AGN populations (see text for more details).}
  \label{fig_mstar_redz}
\end{figure}  

\begin{figure}
\centering
  \includegraphics[width=0.95\columnwidth, height=6.5cm]{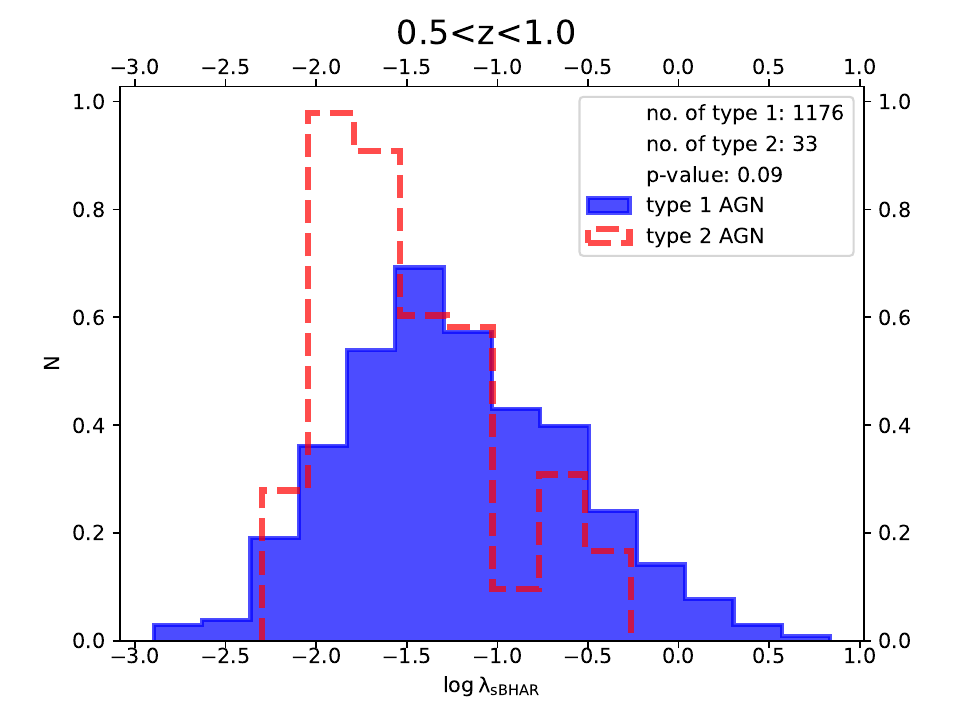}
   \includegraphics[width=0.95\columnwidth, height=6.5cm]{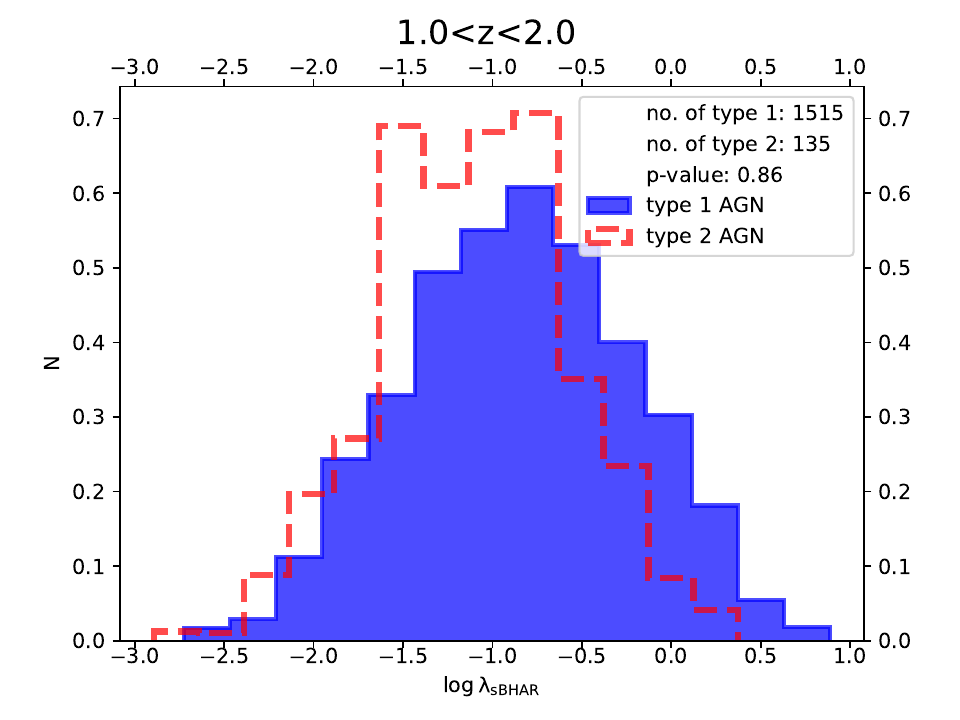}
  \includegraphics[width=0.95\columnwidth, height=6.5cm]{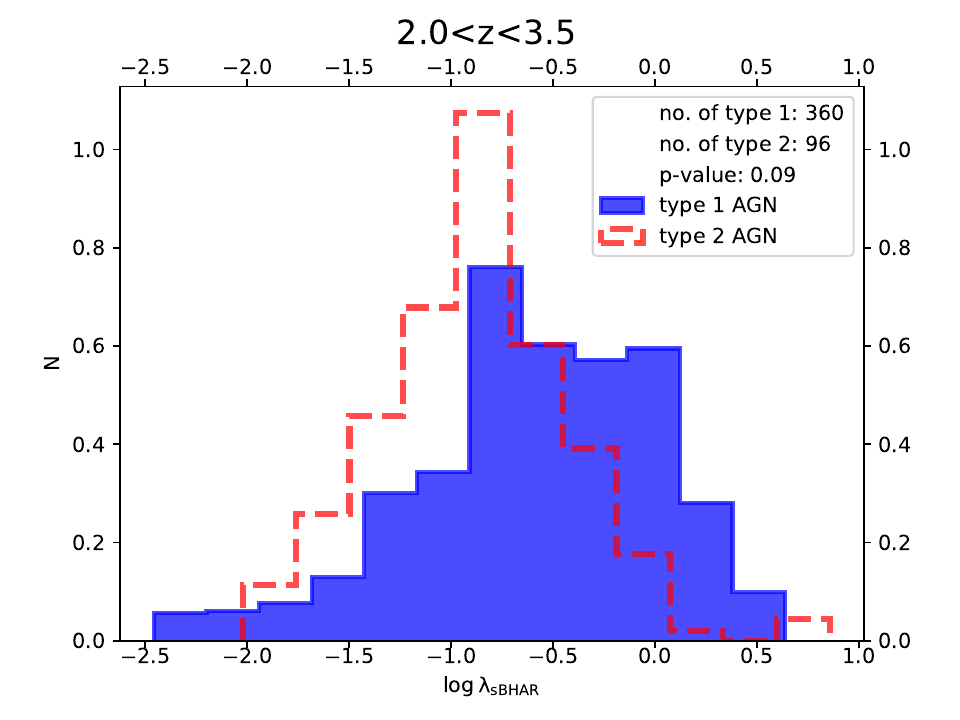}
  \caption{Distributions of $\lambda _{sBHAR}$ of type 1 (blue shaded histogram) and type 2 (red, dashed histogram), at different redshift intervals, as indicated in the title of each panel. Distributions are weighted to account for the different L$_X$ and redshift of the two AGN populations (see text for more details).}
  \label{fig_lambda_redz}
\end{figure}  

\section{Results}
\label{sec_results}

In this section, we perform a comparative analysis of the SFR and M$_*$ between galaxies hosting type 1 and 2 AGN. Additionally, we explore potential distinctions in the $\lambda _{sBHAR}$ for these two AGN populations. 

For that purpose, we split the X-ray AGN dataset into three redshift intervals, that is $\rm 0.5<z<1.0$, $\rm 1.0<z<2.0$ and $\rm 2.0<z<3.5$ and compare the distributions of SFR, M$_*$ and $\lambda _{sBHAR}$ of type 1 and 2 AGN.  In all cases, the distributions are weighted to account for the different redshift and L$_X$ of the two AGN populations, following the process described in, for instance, \cite{Mountrichas2019, Masoura2021, Buat2021, Mountrichas2021b, Koutoulidis2022}.  Specifically, a weight is assigned to each source. This weight is calculated by measuring the joint L$_X$ distributions of the two populations (i.e., we add the number of type 1 and 2 AGN in each L$_X$ bin, in bins of 0.1\,dex) and then normalise the L$_X$ distributions by the total number of sources in each bin. The same procedure is followed for the redshift distributions of the two AGN populations. The total weight that is assigned in each source is the product of the two weights. We make use of these weights in all distributions presented in the remainder of this section.

\subsection{Star-formation rate of type 1 and 2 AGN}

First, we compare the SFR of the two AGN types. The results are shown in Fig. \ref{fig_sfr_redz}, for the three redshift intervals. We notice that, with the exception of the highest redshift interval, type 2 sources tend to have lower SFR compared to their type 1 counterparts. The Kolmogorov-Smirnov (KS) test reveals that this discrepancy has a statistical significance, beyond the level of $2\sigma$ at the lowest redshift bin (p-value of 0.03). In statistical terms, two distributions are considered to differ with a significance of approximately $2\sigma$ for a p-value of 0.05, a threshold commonly employed in similar studies to assess the statistical significance of differences between distributions \citep[e.g.,][]{Zou2019, Mountrichas2021b, Georgantopoulos2023}. The two AGN populations appear to have similar SFR at the highest redshift range probed by the dataset used in our analysis. The (weighted) median SFR values for the two AGN types are presented in Table \ref{table_median}. 

It is worth noting that these trends persist even when we restrict our X-ray dataset to sources with ${\it{specz}}$, resulting in a subset of 1308 type 1 AGN and 150 type 2 AGN within the redshift range of 0.5 < z < 3.5. Additionally, these overarching conclusions hold when we consider the two fields separately, though, the difference in SFR at $\rm 0.5<z<1.0$, between the two AGN types is less pronounced. Finally, we note that the SFR difference is not affected, if we restrict type 1 sources to only those with low levels of polar dust (E$_{B-V}<0.15$, i.e., similarly to type 2 AGN).

\subsection{Stellar mass of type 1 and 2 AGN}

Comparison of the M$_*$ distributions of the two AGN populations, shown in the three panels of Fig. \ref{fig_mstar_redz}, reveals that type 2 sources tend to live in more massive systems compared to type 1, by 0.15-0.30\,dex. The (weighted) median M$_*$ values for the two AGN types are presented in Table \ref{table_median}. This difference although is statistically significant at a level lower than $2\sigma$, appears consistent, across all redshifts spanned by our sample. We also confirm that the results remain unchanged if we restrict the analysis to sources with ${\it{specz}}$ and if we examine the two fields separately.

\subsection{Specific black hole accretion rate of type 1 and 2 AGN}

Finally, we compare the $\lambda _{sBHAR}$ distributions of type 1 and 2 AGN.  $\lambda _{sBHAR}$is defined as:

\begin{equation}
\lambda_{sBHAR}=\rm \frac{k_{bol}\,L_{X,2-10\,keV}}{1.26\times10^{38}\,erg\,s^{-1}\times0.002\frac{M_{*}}{M_\odot}},   
\label{eqn_lambda}
\end{equation}
where $\rm k_{bol}$ is a bolometric correction factor, that converts the $2-10$\,keV X-ray luminosity to AGN bolometric luminosity. We adopt a value of 25 for $\rm k_{bol}$ in line with previous studies \citep[e.g.,][]{Elvis1994, Georgakakis2017, Aird2018, Mountrichas2021c, Mountrichas2022a}, although it is worth noting that lower values \citep[e.g., $\rm k_{bol}=22.4$ in][]{Yang2017} and luminosity-dependent bolometric corrections \citep[e.g.,][]{Hopkins2007a, Lusso2012} have also been employed in the literature.

Although, $\lambda _{sBHAR}$ is often used as a proxy of the Eddington ratio, it is important to acknowledge that the calculation of AGN bolometric luminosity and the inherent scatter in the relation between black hole mass, M$_{BH}$, and M$_*$, may introduce variations in $\lambda_{sBHAR}$ compared to the Eddington ratio, as indicated by previous studies \citep{Lopez2023, Mountrichas2023d}. Nevertheless, our primary objective is to explore potential disparities in the distributions of $\lambda_{sBHAR}$ between type 1 and type 2 AGN. 
 
The $\lambda _{sBHAR}$ distribution of the two AGN classes are presented in the three panels of Fig. \ref{fig_lambda_redz} and the (weighted) median values in Table \ref{table_median}. Based on the results, Type 1 AGN appear to have higher $\lambda _{sBHAR}$ values, by $\sim 0.25$\,dex, in all the three redshift intervals used in our analysis. We confirm that the results are not sensitive to the accuracy of the calculated redshifts (total vs. ${\it{specz}}$ only) and to cosmic variance. 
 
\subsection{The effect of L$_X$}

Subsequently, we split the X-ray dataset into high and low-to-intermediate L$_X$ sources, utilising a cut at $\rm log\,[L_{X,2-10keV}(ergs^{-1})]>44$. Our goal is to examine if the trends we observed are luminosity dependent.
 Table \ref{table_median_lx} displays the weighted median values of SFR, M$_*$, and $\lambda _{sBHAR}$ for both AGN types within the two L$_X$ regimes. Additionally, the table includes the p$-$values derived from conducting the KS-test to assess the distinctions among the different distributions.

In terms of M$_*$ and $\lambda_{sBHAR}$, while the differences appear statistically significant at a level lower than 2$\sigma$, we observe a consistent pattern across all redshifts and luminosity ranges encompassed by our dataset. Specifically, both low-to-intermediate and high L$_X$ type 2 AGN exhibit a preference for more massive systems (albeit with a margin of only $0.1-0.2$ dex) compared to their type 1 counterparts, along with lower values of $\lambda_{sBHAR}$. Therefore, we can conclude that the differences observed in these two parameters (M$_*$ and $\lambda_{sBHAR}$) remain consistent across all redshifts up to 3.5 and are not contingent on AGN luminosity.

Concerning the SFR of the two AGN populations, it appears that type 2 sources with $\rm log,[L_{X,2-10keV}(ergs^{-1})]<44$ tend to have lower SFR compared to type 1 AGN. However, in the case of the most luminous AGN ($\rm log,[L_{X,2-10keV}(ergs^{-1})]>44$), this pattern is only valid at redshifts below 1, and the situation reverses in the highest redshift interval ($\rm z>2$), where luminous type 2 AGN tend to exhibit higher SFR compared to luminous type 1 AGN. Most of these differences appear statistically significant at a $2\sigma$ level.

Overall, we conclude that type 2 sources prefer to live in more massive host galaxies and tend to have lower $\lambda_{sBHAR}$ compared to type 1 X-ray AGN, at all redshifts and L$_X$ spanned by our dataset. Moreover, low-to-moderate L$_X$ type 2 systems appear to have lower SFR compared to their type 1 counterparts. This picture reverses at high L$_X$ ($\rm log,[L_{X,2-10keV}(ergs^{-1})]>44$) and redshift $\rm z>2$. The fact that most of these differences are statistically significant at a lever lower than 2$\sigma$ can be attributed to the contamination in CIGALE's classification. As noted, previous studies have shown that CIGALE misclassifies $10-20\%$ of the sources. 

\subsection{Classification based on X-ray obscuration}

To facilitate a more direct comparison with previous studies that relied on X-ray criteria to classify AGN, we divide the 3\,312 sources with reliable CIGALE classification (Table \ref{table_classified}) into X-ray absorbed and unabsorbed, using a cut at $\rm N_H=10^{23}$\,cm$^{-2}$ and repeat our analysis. There are 145 X-ray absorbed and 3167 X-ray unabsorbed AGN in our dataset. The results are presented in Table \ref{table_nh}. Similarly to the results obtained using the classification from CIGALE, absorbed AGN tend to have lower $\lambda_{sBHAR}$ and live, on average, in more massive galaxies that exhibit lower SFR (although, the latter is now observed at all redshifts spanned by our sample) compared to their unabsorbed counterparts. However, with the exception of the SFR at $\rm z<2$, all other differences do not appear statistically significant ($<2\,\sigma$). 

Since the same trends are found independent of redshift, we then merge the three redshift bins and split the sources into low and high L$_X$, utilizing a threshold at $\rm log[L_{X,2-10keV}(ergs^{-1})]=44$). The results are displayed in Table \ref{table_nh_lx}. Based on our findings, it is primarily the low-to-intermediate L$_X$ AGN that present differences in the host galaxy and SMBH properties of the two AGN populations. These differences also appear statistically significant at a level of $\approx 2\sigma$. We also note, that the observed trends diminish when we lower the $\rm N_H$ threshold used to classify AGN to $10^{22}$\,cm$^{-2}$.

\begin{table*}
\caption{Weighted median values of SFR, M$_*$ and $\lambda _{sBHAR}$ for X-ray absorbed and unabsorbed sources, utilizing a threshold of $\rm log\,[N_H\,(cm^{-2})]=23$, in the three redshift intervals used in our analysis. The number of sources and the p$-$values yielded by applying KS$-$tests, are also presented.}
\centering
\setlength{\tabcolsep}{3mm}
\begin{tabular}{ccccccc}
 \hline
 & \multicolumn{2}{c}{0.5<z<1.0} &  \multicolumn{2}{c}{1.0<z<2.0} &  \multicolumn{2}{c}{2.0<z<3.5}\\
 \hline
$\rm log\,[N_H\,(cm^{-2})]$   & $<10^{23}$ & $>10^{23}$ & $<10^{23}$ & $>10^{23}$  & $<10^{23}$ & $>10^{23}$ \\
 \hline
number of sources & 1\,631 & 17 & 1\,130 & 80 & 406 & 48  \\ 
 \hline
log\,SFR  &  1.65 & 0.07 & 2.20 & 1.33 & 2.44 & 1,89 \\
p$-$value (SFR) & \multicolumn{2}{c}{0.03} & \multicolumn{2}{c}{0.03} & \multicolumn{2}{c}{0.31} \\
 \hline
log\,M$_*$ & 10.89 & 10.96 & 11.05 & 11.11 & 11.27 & 11.33 \\
p$-$value (M$_*$) & \multicolumn{2}{c}{0.90} & \multicolumn{2}{c}{0.90} & \multicolumn{2}{c}{0.31} \\
 \hline
log\,$\lambda _{sBHAR}$ & -1.27 & -1.48 & -0.89 & -0.98 & -0.76 & -0.88 \\
p$-$vaule ($\lambda _{sBHAR}$) & \multicolumn{2}{c}{0.22} & \multicolumn{2}{c}{0.87} & \multicolumn{2}{c}{0.23}\\
  \hline
\label{table_nh}
\end{tabular}
\end{table*}

\begin{table*}
\caption{Same as in Table \ref{table_nh}, but now splitting the X-ray dataset into two L$_X$ bins, utilizing a threshold at $\rm log\,[L_{X,2-10keV}(ergs^{-1})]=44$.}
\centering
\setlength{\tabcolsep}{5mm}
\begin{tabular}{ccccc}
 \hline
 & \multicolumn{4}{c}{0.5<z<3.5} \\
 \hline
 & \multicolumn{2}{c}{log\,L$_X<44$} & \multicolumn{2}{c}{log\,L$_X>44$} \\

$\rm log\,[N_H\,(cm^{-2})]$   & $<10^{23}$ & $>10^{23}$  & $<10^{23}$ & $>10^{23}$\\
 \hline
number of sources & 1392 & 69 & 1775  & 76 \\ 
 \hline
log\,SFR  &  1.69 & 1.06 &  2.39 & 1.49 \\
p$-$value (SFR) & \multicolumn{2}{c}{0.02} & \multicolumn{2}{c}{0.36}\\
 \hline
log\,M$_*$ & 11.03 & 11.27  &  11.07 & 11.04 \\
p$-$value (M$_*$) & \multicolumn{2}{c}{0.08}& \multicolumn{2}{c}{0.78} \\
 \hline
log\,$\lambda _{sBHAR}$ & -1.38 & -1.71 &  -0.66 &  -0.62 \\
p$-$vaule ($\lambda _{sBHAR}$) & \multicolumn{2}{c}{0.05} & \multicolumn{2}{c}{0.84}\\
  \hline
\label{table_nh_lx}
\end{tabular}
\end{table*}

\section{Discussion}
\label{sec_discussion}

In these section, we discuss how our results compare with the findings of prior studies. Additionally, we delve into the influence of varying classification criteria on the reported outcomes.

\subsection{Obscuration and M$_*$}
\cite{Zou2019} divided X-ray sources in the COSMOS field into type 1 and 2, based on their optical spectra, morphologies and variability. They found that type 2 sources are inclined to inhabit more massive systems compared to type 1, by $0.1-0.2$\,dex, up to $\rm z \approx 3.5$. \cite{Mountrichas2021b} examined X-ray detected AGN in the XMM-XXL field, at a median $\rm z\approx 0.5$ and classified AGN into two types using the classification that is available in the XXL catalogue \citep{Menzel2016} and is based on optical spectra. Their results are similar to those reported by \cite{Zou2019}. Our findings align with the results of these previous studies. Different M$_*$ for the different AGN populations have also been reported by studies that used X-ray classifcation criteria (\citealt{Lanzuisi2017, Georgantopoulos2023}, but see \citealt{Masoura2021, Mountrichas2021b}).

\subsection{Obscuration and SFR}

Both of the previously mentioned investigations \citep{Zou2019, Mountrichas2021b} do not discern a significant disparity in the SFR of host galaxies between the two AGN populations. Our analysis, however, suggests that the hosts of type 1 and 2 sources have different SFR and this difference presents a dependence on redshift and L$_X$. It is worth noting that the sources utilized in \cite{Mountrichas2021b} primary probe lower redshifts and span a narrower L$_X$ range compared to our dataset, as indicated in their Fig. 7. To facilitate a better comparison with \cite{Zou2019}, we conduct a supplementary analysis by limiting our sample to sources within the COSMOS field. The outcomes of this restricted analysis indicate that the two AGN populations exhibit smaller differences regarding their SFR distributions, up to a redshift of $\rm z = 2$ and the statistical significance of these differences are notably low (p$-$value $\sim 0.60$). In the highest redshift interval ($\rm 2.0 < z < 3.5$), the results obtained are similar to those using both datasets 

Difference in the SFR has also been reported in the literature by studies that classified X-ray AGN using X-ray absorption criteria (\citealt{Georgantopoulos2023}, but see \citealt{Lanzuisi2017, Masoura2021, Mountrichas2021b}). \cite{Georgantopoulos2023} used AGN in the COSMOS field and a high $\rm N_H$ threshold ($\rm N_H=10^{23}$\,cm$^{-2}$) to classify X-ray sources into absorbed and unabsorbed. They observed that absorbed sources exhibited a tendency towards lower levels of specific star formation rate (sSFR, defined as $\rm sSFR=\frac{SFR}{M*}$) when compared to unabsorbed sources. Georgantopoulos et al. attributed the disparities observed in their results, in contrast to prior research that found no discernible distinctions in the properties of host galaxies between X-ray absorbed and unabsorbed AGN \citep[e.g.,][]{Masoura2021, Mountrichas2021c}, or research employing optical criteria \citep[e.g.,][]{Zou2019, Mountrichas2021b}, to variances in the $\rm N_H$ thresholds employed and/or the range of luminosities investigated by different sample sets. Mountrichas et al. (submitted), used X-ray AGN from the 4XMM dataset and applied strict X-ray criteria for the AGN classification ($\rm N_H=10^{23}$\,cm$^{-2}$), taking also into account the uncertainties associated with the $\rm N_H$ measurements. Their findings closely mirrored those reported in the study by \cite{Georgantopoulos2023} concerning the host galaxy properties of the two AGN populations. The aforementioned discrepancies were attributed to the varying ranges of L$_X$ that different studies examined and not on the $\rm N_H$ threshold utilized for the AGN classification. Specifically, their analysis demonstrated that the observed distinctions in host galaxy properties between obscured and unobscured AGN tended to diminish notably at higher levels of X-ray luminosity (specifically, when $\rm log[L_{X,2-10keV}(ergs^{-1})]>44$).

\subsection{Obscuration and Eddington ratio}

Our study suggest that type 2 AGN tend to have lower $\lambda_{sBHAR}$ compared to type 1. Prior works that categorised AGN based on optical criteria did not examine their difference in n$_{Edd}$ or $\lambda_{sBHAR}$. However, differences in n$_{Edd}$ between the two populations have been reported by studies that used X-criteria to classify AGN. These studies have shown that X-ray absorbed AGN exhibit, on average, lower n$_{Edd}$ compared to their X-ray unabsorbed counterparts \citep{Ricci2017, Ananna2022, Georgantopoulos2023}, in agreement with our findings.

\subsection{Concluding remarks}

The picture that emerges is that when AGN are classified as type 1 and 2 (e.g., using optical criteria), then type 2 AGN appear to inhabit, on average, galaxies with higher M$_*$ than type 1 \cite[this study,][]{Zou2019, Mountrichas2021b}. Type 2 sources are also hosted by systems exhibiting lower SFR, but this difference is not universal and seems to hinge on redshift and, most importantly, on the L$_X$ probed. Furthermore, type 2 AGN also tend to have lower $\lambda_{sBHAR}$ (or n$_{Edd}$).

\begin{figure}
\centering
  \includegraphics[width=0.95\columnwidth, height=7.2cm]{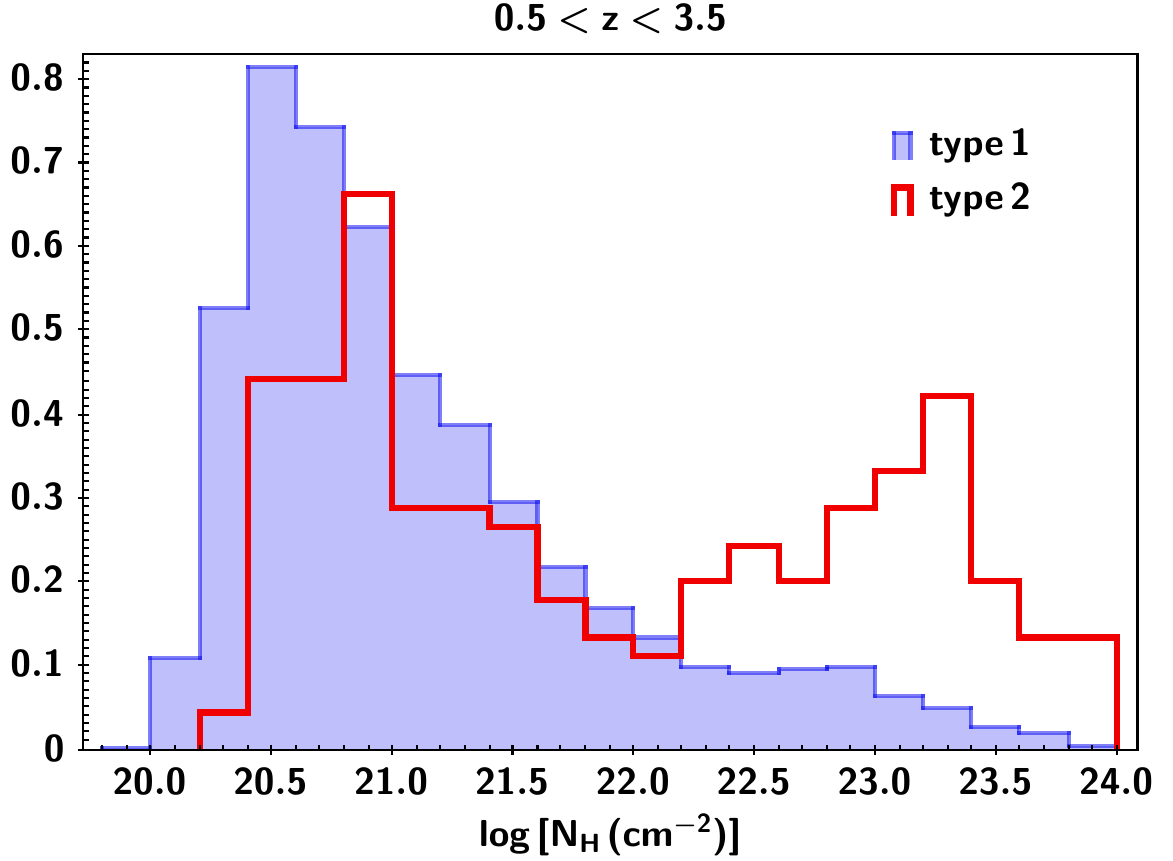}   
  \caption{Distribution of $\rm N_H$, for type 1 (blue, shaded histogram) and type 2 (red, dashed histogram) AGN.}
  \label{fig_nh}
\end{figure}

It is important to highlight that comparing results between studies that employ optical and X-ray criteria for AGN classification can be challenging due to the considerable scatter in the correlation between optical obscuration and X-ray absorption. For instance, a source may be heavily X-ray obscured with broad UV/optical lines \citep[e.g.,][]{Merloni2014, Li2019} and can be optically classified as type 2, without being necessarily X-ray absorbed \citep[e.g.,][]{Masoura2020}. Fig. \ref{fig_nh} presents the distribution of $\rm N_H$ for the type 1 and type 2 sources used in our analysis (at $\rm 0.5<z<3.5$). While type 2 AGN typically exhibit higher $\rm N_H$ values than type 1 AGN (median $\rm log\,[N_H\,(cm^{-2})]=21.73$ and 20.90, for type 2 and type 1, respectively), it is crucial to acknowledge that these criteria would result in the classification of a significant number of sources in different ways. It is important to highlight that CIGALE lacks sensitivity to incremental changes in the inclination angle. Consequently, intermediate values of the estimated inclination angle should not be deemed reliable for categorizing AGN into sub-classes. Therefore, certain sources identified as type 2 by CIGALE might actually be more akin to type 1.8/1.9, exhibiting similarities with type 1 rather than type 2 sources, as exemplified in studies such as \cite{Trippe2010, HernandezGarcia2017}. These sources may also be characterized as X-ray unabsorbed, as indicated by \cite{Shimizu2018}. It is noteworthy, though, that these observations are not indicative of misclassification by CIGALE. Similar results were reported in \cite{Mountrichas2021b}, where the categorization of AGN in type 1 and 2 was based on optical spectra (see the right panel of their Fig. 7). Further investigation reveals, that type 1 AGN with N$_H>10^{22}$\,cm$^{-2}$ have a tendency for increased levels of polar dust, based on CIGALE measurements compared to type 1 AGN with lower $\rm N_H$. Specifically, the median $E_{B-V}$ value for the former class is 0.22 and for the latter 0.15. \cite{Mountrichas2021b} found that the majority of sources ($\sim 80\%$, i.e., $\frac{105}{105+27}$) classified as type 1 by CIGALE that also present elevated levels of polar dust ($E_{B-V}>0.15$) are spectroscopically confirmed as type 1 (see their Fig. 8).  Furthermore, there exists a significant fraction  ($\sim 50\%$) of type 2 AGN characterised by low levels of X-ray obscuration ($\rm log\,[N_H\,(cm^{-2})]<22$). As previously mentioned, such sources have been reported in the literature \citep[e.g.,][]{Masoura2020}. These findings collectively underscore the intricate distribution of gas and dust, contributing to the diverse array of AGN properties \citep[e.g.][]{Lyu2018, EsparzaArredondo2021}.

\section{Conclusions}
\label{sec_conclusions}

In this work, we used X-ray AGN detected in the eFEDS and {\it{COSMOS-Legacy}} fields to study the SMBH and host galaxy properties of type 1 and 2 AGN. The sources were classified based on the results of the SED fitting analysis, using the CIGALE code. To ensure the robustness of our analysis, we applied stringent selection criteria, ensuring that only sources with reliable host galaxy properties and classifications were included. Consequently, our final dataset consisted of 3\,312 sources, with approximately 85\% of them located in the eFEDS field. Out of these sources,  3\,049 are type 1 AGN and 263 are type 2 AGN.  We note, that according to previous studies, CIGALE's classification performance has a success rate of about $\sim 80-85\%$. 
The primary findings of our study can be summarized as follows:

\begin{itemize}

\item[$\bullet$] Type 2 AGN tend to inhabit in more massive systems, by $0.2-0.3$\,dex (in logarithmic scale), compared to their type 1 counterparts. Their specific black hole accretion rate, a proxy of the Eddington ratio, is, on average, lower in the case of type 2 sources compared to type 1, by $\sim 0.3$\,dex in logarithmic scales. These differences although appear to have a statistical significance lower than $2\sigma$, they are observed across all redshifts and X-ray luminosities probed by our dataset ($\rm 0.5<z<3.5$, $\rm 42<log,[L_{X,2-10keV}(ergs^{-1})]<46$).

\item[$\bullet$] Type 2 AGN tend to have lower SFR compared to type 1 AGN at $\rm z<1$. Conversely, this picture reverses at high redshifts ($\rm z>2$) and X-ray luminosities ($\rm log,[L_{X,2-10keV}(ergs^{-1})]>44$). These differences are statistically significant at approximately a 2$\sigma$ confidence level.

\item[$\bullet$] Similar trends are discernible  when we classify the 3\,312 AGN based on their X-ray obscuration, applying a $\rm N_H$ cut at $10^{23}$\,cm$^{-2}$. However, it is noteworthy that these observed differences are pronounced for non-luminous AGN ($\rm log[L_{X,2-10keV}(ergs^{-1})]<44$) and also tend to diminish when we lower the $\rm N_H$ threshold used for AGN classification.

\end{itemize}

The results from our analysis suggest that, irrespective of whether we employ optical or X-ray criteria to categorize AGN as obscured or unobscured, the disparities in the host galaxy and SMBH properties between the two AGN populations exhibit similar trends. However, these differences are sensitive to the L$_X$ regime probed and the stringency of the applied X-ray criteria. Thus, caution has to be taken  when we compare results from different studies.

\begin{acknowledgements}
This project has received funding from the European Union's Horizon 2020 research and innovation program under grant agreement no. 101004168, the XMM2ATHENA project.
This research has made use of TOPCAT version 4.8 \citep{Taylor2005} and Astropy \citep{PriceWhelan2022}.

\end{acknowledgements}

\bibliography{mybib}

\begin{thebibliography}{79}
\expandafter\ifx\csname natexlab\endcsname\relax\def\natexlab#1{#1}\fi

\bibitem[{Aird {et~al.}(2018)Aird, Coil, \& Georgakakis}]{Aird2018}
Aird, J., Coil, A.~L., \& Georgakakis, A. 2018, Monthly Notices of the Royal
  Astronomical Society, 474, 1225

\bibitem[{Ananna {et~al.}(2022)Ananna, Urry, Ricci, Natarajan, Hickox,
  Trakhtenbrot, Treister, Weigel, Ueda, Koss, Bauer, Temple, Balokovi{\'{c}},
  Mushotzky, Auge, Sanders, Kakkad, Sartori, Marchesi, Harrison, Stern, Oh,
  Caglar, Powell, Podjed, \& Mej{\'{\i}}a-Restrepo}]{Ananna2022}
Ananna, T.~T., Urry, C.~M., Ricci, C., {et~al.} 2022, The Astrophysical Journal
  Letters, 939, L13

\bibitem[{{Arnouts} {et~al.}(1999){Arnouts}, {Cristiani}, {Moscardini},
  {Matarrese}, {Lucchin}, {Fontana}, \& {Giallongo}}]{Arnouts1999}
{Arnouts}, S., {Cristiani}, S., {Moscardini}, L., {et~al.} 1999, MNRAS, 310,
  540

\bibitem[{Baldry {et~al.}(2018)Baldry, Liske, Brown, Robotham, Driver, Dunne,
  Alpaslan, Brough, Cluver, Eardley, Farrow, Heymans, Hildebrandt, Hopkins,
  Kelvin, Loveday, Moffett, Norberg, Owers, Taylor, Wright, Bamford,
  Bland-Hawthorn, Bourne, Bremer, Colless, Conselice, Croom, Davies, Foster,
  Grootes, Holwerda, Jones, Kafle, Kuijken, Lara-Lopez,
  L{\'{o}}pez-S{\'{a}}nchez, Meyer, Phillipps, Sutherland, van Kampen, \&
  Wilkins}]{Baldry2018}
Baldry, I.~K., Liske, J., Brown, M. J.~I., {et~al.} 2018, Monthly Notices of
  the Royal Astronomical Society, 474, 3875

\bibitem[{Blanton {et~al.}(2017)Blanton, Bershady, Abolfathi, Albareti, Prieto,
  Almeida, Alonso-García, Anders, Anderson, Andrews, Aquino-Ortíz,
  Aragón-Salamanca, Argudo-Fernández, Armengaud, Aubourg, Avila-Reese,
  Badenes, Bailey, Barger, Barrera-Ballesteros, Bartosz, Bates, Baumgarten,
  Bautista, Beaton, Beers, Belfiore, Bender, Berlind, Bernardi, Beutler, Bird,
  Bizyaev, Blanc, Blomqvist, Bolton, Boquien, Borissova, van~den Bosch, Bovy,
  Brandt, Brinkmann, Brownstein, Bundy, Burgasser, Burtin, Busca, Cappellari,
  Carigi, Carlberg, Rosell, Carrera, Cherinka, Cheung, Chew, Chiappini, Choi,
  Chojnowski, Chuang, Chung, Cirolini, Clerc, Cohen, Comparat, da~Costa,
  Cousinou, Covey, Crane, Croft, Cruz-Gonzalez, Cuadra, Cunha, Damke, Darling,
  Davies, Dawson, de~la Macorra, Lee, Delubac, Mille, Diamond-Stanic,
  Cano-Díaz, Donor, Downes, Drory, du~Mas~des Bourboux, Duckworth, Dwelly,
  Dyer, Ebelke, Eisenstein, Emsellem, Eracleous, Escoffier, Evans, Fan,
  Fernández-Alvar, Fernandez-Trincado, Feuillet, Finoguenov, Fleming,
  Font-Ribera, Fredrickson, Freischlad, Frinchaboy, Galbany, Garcia-Dias,
  García-Hernández, Gaulme, Geisler, Gelfand, Gil-Marín, Gillespie, Goddard,
  Gonzalez-Perez, Grabowski, Green, Grier, Gunn, Guo, Guy, Hagen, Hahn, Hall,
  Harding, Hasselquist, Hawley, Hearty, Hernández, Ho, Hogg,
  Holley-Bockelmann, Holtzman, Holzer, Huehnerhoff, Hutchinson, Hwang,
  Ibarra-Medel, da~Silva~Ilha, Ivans, Ivory, Jackson, Jensen, Johnson, Jones,
  Jönsson, Jullo, Kamble, Kinemuchi, Kirkby, Kitaura, Klaene, Knapp, Kneib,
  Kollmeier, Lacerna, Lane, Lang, Law, Lazarz, Goff, Liang, Li, LI, Lima, Lin,
  Lin, de~Lis, Liu, de~Icaza~Lizaola, Long, Lucatello, Lundgren, MacDonald,
  Machado, MacLeod, Mahadevan, Maia, Maiolino, Majewski, Malanushenko,
  Malanushenko, Manchado, Mao, Maraston, Marques-Chaves, Masters, McBride,
  McDermid, McGrath, McGreer, Peña, Melendez, Merloni, Merrifield, Meszaros,
  Meza, Minchev, Minniti, Miyaji, More, Mulchaey, Müller-Sánchez, Muna,
  Munoz, Myers, Nair, Nandra, do~Nascimento, Negrete, Ness, Newman, Nichol,
  Nidever, Nitschelm, Ntelis, O'Connell, Oelkers, Oravetz, Oravetz, Pace,
  Padilla, Palanque-Delabrouille, Palicio, Pan, Parikh, Pâris, Park, Patten,
  Peirani, Pellejero-Ibanez, Penny, Percival, Perez-Fournon, Petitjean, Pieri,
  Pinsonneault, Pisani, Poleski, Prada, Prakash, de~Andrade~Queiroz, Raddick,
  Raichoor, Rembold, Richstein, Riffel, Riffel, Rix, Robin, Rockosi,
  Rodríguez-Torres, Roman-Lopes, Román-Zúñiga, Rosado, Ross, Rossi, Ruan,
  Ruggeri, Rykoff, Salazar-Albornoz, Salvato, Sánchez, Aguado,
  Sánchez-Gallego, Santana, Santiago, Sayres, Schiavon, da~Silva~Schimoia,
  Schlafly, Schlegel, Schneider, Schultheis, Schuster, Schwope, Seo, Shao,
  Shen, Shetrone, Shull, Simon, Skinner, Skrutskie, Slosar, Smith, Sobeck,
  Sobreira, Somers, Souto, Stark, Stassun, Stauffer, Steinmetz,
  Storchi-Bergmann, Streblyanska, Stringfellow, Suárez, Sun, Suzuki, Szigeti,
  Taghizadeh-Popp, Tang, Tao, Tayar, Tembe, Teske, Thakar, Thomas, Thompson,
  Tinker, Tissera, Tojeiro, Toledo, de~la Torre, Tremonti, Troup, Valenzuela,
  Valpuesta, Vargas-González, Vargas-Magaña, Vazquez, Villanova, Vivek, Vogt,
  Wake, Walterbos, Wang, Weaver, Weijmans, Weinberg, Westfall, Whelan, Wild,
  Wilson, Wood-Vasey, Wylezalek, Xiao, Yan, Yang, Ybarra, Yèche, Zakamska,
  Zamora, Zarrouk, Zasowski, Zhang, Zhao, Zheng, Zhou, Zhu, Zoccali, \&
  Zou}]{Blanton2017}
Blanton, M.~R., Bershady, M.~A., Abolfathi, B., {et~al.} 2017, AJ
  [\eprint{1703.00052}], 35

\bibitem[{Boquien {et~al.}(2019)Boquien, Burgarella, Roehlly, Buat, Ciesla,
  Corre, Inoue, \& Salas}]{Boquien2019}
Boquien, M., Burgarella, D., Roehlly, Y., {et~al.} 2019, Astronomy {\&}
  Astrophysics, 622, A103

\bibitem[{Brunner {et~al.}(2022)Brunner, Liu, Lamer, Georgakakis, Merloni,
  Brusa, Bulbul, Dennerl, Friedrich, Liu, Maitra, Nandra, Ramos-Ceja, Sanders,
  Stewart, Boller, Buchner, Clerc, Comparat, Dwelly, Eckert, Finoguenov,
  Freyberg, Ghirardini, Gueguen, Haberl, Kreykenbohm, Krumpe, Osterhage,
  Pacaud, Predehl, Reiprich, Robrade, Salvato, Santangelo, Schrabback, Schwope,
  \& Wilms}]{Brunner2022}
Brunner, H., Liu, T., Lamer, G., {et~al.} 2022, Astronomy \& Astrophysics, 661,
  A1

\bibitem[{Bruzual \& Charlot(2003)}]{Bruzual_Charlot2003}
Bruzual, G. \& Charlot, S. 2003, MNRAS, 344, 1000

\bibitem[{Buat {et~al.}(2019)Buat, Ciesla, Boquien, Ma{\l}ek, \&
  Burgarella}]{Buat2019}
Buat, V., Ciesla, L., Boquien, M., Ma{\l}ek, K., \& Burgarella, D. 2019,
  Astronomy {\&} Astrophysics, 632, A79

\bibitem[{Buat {et~al.}(2021)Buat, Mountrichas, Yang, Boquien, Roehlly,
  Burgarella, Stalevski, Ciesla, \& Theul{\'{e}}}]{Buat2021}
Buat, V., Mountrichas, G., Yang, G., {et~al.} 2021, A\&A, 654, A93

\bibitem[{Charlot \& Fall(2000)}]{Charlot_Fall_2000}
Charlot, S. \& Fall, S.~M. 2000, ApJ, 539, 718

\bibitem[{Ciotti \& Ostriker(1997)}]{Ciotti1997}
Ciotti, L. \& Ostriker, J.~P. 1997, The Astrophysical Journal, 487, L105

\bibitem[{Civano {et~al.}(2016)Civano, Marchesi, Comastri, Urry, Elvis,
  Cappelluti, Puccetti, Brusa, Zamorani, Hasinger, Aldcroft, Alexander,
  Allevato, Brunner, Capak, Finoguenov, Fiore, Fruscione, Gilli, Glotfelty,
  Griffiths, Hao, Harrison, Jahnke, Kartaltepe, Karim, LaMassa, Lanzuisi,
  Miyaji, Ranalli, Salvato, Sargent, Scoville, Schawinski, Schinnerer,
  Silverman, Smolcic, Stern, Toft, Trakhenbrot, Treister, \&
  Vignali}]{Civano2016}
Civano, F., Marchesi, S., Comastri, A., {et~al.} 2016, ApJ, 819, 62

\bibitem[{{Dale} {et~al.}(2014){Dale}, {Helou}, {Magdis}, {Armus},
  {D{\'{\i}}az-Santos}, \& {Shi}}]{Dale2014}
{Dale}, D.~A., {Helou}, G., {Magdis}, G.~E., {et~al.} 2014, ApJ, 784, 83

\bibitem[{Drinkwater {et~al.}(2018)Drinkwater, Byrne, Blake, Glazebrook,
  Brough, Colless, Couch, Croton, Croom, Davis, Forster, Gilbank, Hinton,
  Jelliffe, Jurek, hui Li, Martin, Pimbblet, Poole, Pracy, Sharp, Smillie,
  Spolaor, Wisnioski, Woods, Wyder, \& Yee}]{Drinkwater2018}
Drinkwater, M.~J., Byrne, Z.~J., Blake, C., {et~al.} 2018, Monthly Notices of
  the Royal Astronomical Society, 474, 4151

\bibitem[{Elvis {et~al.}(1994)Elvis, Wilkes, McDowell, Green, Bechtold,
  Willner, Oey, Polomski, \& Cutri}]{Elvis1994}
Elvis, M., Wilkes, B.~J., McDowell, J.~C., {et~al.} 1994, The Astrophysical
  Journal Supplement Series, 95, 1

\bibitem[{Esparza-Arredondo {et~al.}(2021)Esparza-Arredondo,
  Gonzalez-Mart{\'{\i}}n, Dultzin, Masegosa, Ramos-Almeida,
  Garc{\'{\i}}a-Bernete, Fritz, \& Osorio-Clavijo}]{EsparzaArredondo2021}
Esparza-Arredondo, D., Gonzalez-Mart{\'{\i}}n, O., Dultzin, D., {et~al.} 2021,
  Astronomy \& Astrophysics, 651, A91

\bibitem[{Georgakakis {et~al.}(2017)Georgakakis, Aird, Schulze, Dwelly,
  Salvato, Nandra, Merloni, \& Schneider}]{Georgakakis2017}
Georgakakis, A., Aird, J., Schulze, A., {et~al.} 2017, MNRAS, 471, 1976

\bibitem[{Georgantopoulos {et~al.}(2023)Georgantopoulos, Pouliasis,
  Mountrichas, der Wel, Marchesi, \& Lanzuisi}]{Georgantopoulos2023}
Georgantopoulos, I., Pouliasis, E., Mountrichas, G., {et~al.} 2023, Astronomy
  \& Astrophysics, 673, A67

\bibitem[{Hernández-García {et~al.}(2017)Hernández-García, Masegosa,
  González-Martín, Márquez, Guainazzi, \& Panessa}]{HernandezGarcia2017}
Hernández-García, L., Masegosa, J., González-Martín, O., {et~al.} 2017,
  Astronomy \& Astrophysics, 602, A65

\bibitem[{Hopkins {et~al.}(2006)Hopkins, Hernquist, Cox, Matteo, Robertson, \&
  Springel}]{Hopkins2006}
Hopkins, P.~F., Hernquist, L., Cox, T.~J., {et~al.} 2006, The Astrophysical
  Journal Supplement Series, 163, 1

\bibitem[{Hopkins {et~al.}(2007)Hopkins, Richards, \& Hernquist}]{Hopkins2007a}
Hopkins, P.~F., Richards, G.~T., \& Hernquist, L. 2007, The Astrophysical
  Journal, 654, 731

\bibitem[{{Ilbert} {et~al.}(2006)}]{Ilbert2006}
{Ilbert}, O. {et~al.} 2006, A\&A, 457, 841

\bibitem[{Koutoulidis {et~al.}(2022)Koutoulidis, Mountrichas, Georgantopoulos,
  Pouliasis, \& Plionis}]{Koutoulidis2022}
Koutoulidis, L., Mountrichas, G., Georgantopoulos, I., Pouliasis, E., \&
  Plionis, M. 2022, Astronomy {\&} Astrophysics, 658, A35

\bibitem[{Kuijken {et~al.}(2019)Kuijken, Heymans, Dvornik, Hildebrandt,
  de~Jong, Wright, Erben, Bilicki, Giblin, Shan, Getman, Grado, Hoekstra,
  Miller, Napolitano, Paolilo, Radovich, Schneider, Sutherland, Tewes, Tortora,
  Valentijn, \& Kleijn}]{Kuijken2019}
Kuijken, K., Heymans, C., Dvornik, A., {et~al.} 2019, Astronomy {\&}
  Astrophysics, 625, A2

\bibitem[{Kuijken {et~al.}(2015)Kuijken, Heymans, Hildebrandt, Nakajima, Erben,
  de~Jong, Viola, Choi, Hoekstra, Miller, van Uitert, Amon, Blake, Brouwer,
  Buddendiek, Conti, Eriksen, Grado, Harnois-D{\'{e}}raps, Helmich, Herbonnet,
  Irisarri, Kitching, Klaes, Barbera, Napolitano, Radovich, Schneider,
  Sif{\'{o}}n, Sikkema, Simon, Tudorica, Valentijn, Kleijn, \& van
  Waerbeke}]{Kuijken2015}
Kuijken, K., Heymans, C., Hildebrandt, H., {et~al.} 2015, Monthly Notices of
  the Royal Astronomical Society, 454, 3500

\bibitem[{Laigle {et~al.}(2016)Laigle, McCracken, Ilbert, Hsieh, Davidzon,
  Capak, Hasinger, Silverman, Pichon, Coupon, Aussel, Borgne, Caputi, Cassata,
  Chang, Civano, Dunlop, Fynbo, Kartaltepe, Koekemoer, F{\`{e}}vre, Floc'h,
  Leauthaud, Lilly, Lin, Marchesi, Milvang-Jensen, Salvato, Sanders, Scoville,
  Smolcic, Stockmann, Taniguchi, Tasca, Toft, Vaccari, \& Zabl}]{Laigle2016}
Laigle, C., McCracken, H.~J., Ilbert, O., {et~al.} 2016, ApJS, 224, 24

\bibitem[{{Lanzuisi} {et~al.}(2017)}]{Lanzuisi2017}
{Lanzuisi}, G. {et~al.} 2017, A\&A, 602, 13

\bibitem[{Li {et~al.}(2019)Li, Xue, Sun, Liu, Vito, Brandt, Hughes, Yang,
  Tozzi, Zhu, Zheng, Luo, Chen, Vignali, Gilli, \& Shu}]{Li2019}
Li, J., Xue, Y., Sun, M., {et~al.} 2019, The Astrophysical Journal, 877, 5

\bibitem[{Liu {et~al.}(2022)Liu, Buchner, Nandra, Merloni, Dwelly, Sanders,
  Salvato, Arcodia, Brusa, Wolf, Georgakakis, Boller, Krumpe, Lamer, Waddell,
  Urrutia, Schwope, Robrade, Wilms, Dauser, Comparat, Toba, Ichikawa, Iwasawa,
  Shen, \& Medel}]{Liu2022}
Liu, T., Buchner, J., Nandra, K., {et~al.} 2022, Astronomy \& Astrophysics,
  661, A5

\bibitem[{Lopez {et~al.}(2023)Lopez, Brusa, Bonoli, Shankar, Acharya, Laloux,
  Dolag, Georgakakis, \& Lapi}]{Lopez2023}
Lopez, I.~E., Brusa, M., Bonoli, S., {et~al.} 2023, Astronomy {\&}
  Astrophysics, 672, A137

\bibitem[{{Lusso} {et~al.}(2012)}]{Lusso2012}
{Lusso}, E. {et~al.} 2012, MNRAS, 425, 623

\bibitem[{Lyu \& Rieke(2018)}]{Lyu2018}
Lyu, J. \& Rieke, G.~H. 2018, The Astrophysical Journal, 866, 92

\bibitem[{Ma{\l}ek {et~al.}(2018)Ma{\l}ek, Buat, Roehlly, Burgarella, Hurley,
  Shirley, Duncan, Efstathiou, Papadopoulos, Vaccari, Farrah, Marchetti, \&
  Oliver}]{Malek2018}
Ma{\l}ek, K., Buat, V., Roehlly, Y., {et~al.} 2018, Astronomy {\&}
  Astrophysics, 620, A50

\bibitem[{Marchesi {et~al.}(2016)Marchesi, Civano, Elvis, Salvato, Brusa,
  Comastri, Gilli, Hasinger, Lanzuisi, Miyaji, Treister, Urry, Vignali,
  Zamorani, Allevato, Cappelluti, Cardamone, Finoguenov, Griffiths, Karim,
  Laigle, LaMassa, Jahnke, Ranalli, Schawinski, Schinnerer, Silverman, Smolcic,
  Suh, \& Trakhtenbrot}]{Marchesi2016}
Marchesi, S., Civano, F., Elvis, M., {et~al.} 2016, ApJ, 817, 34

\bibitem[{Masoura {et~al.}(2020)Masoura, Georgantopoulos, Mountrichas, Vignali,
  Koulouridis, Chiappetti, Fotopoulou, Paltani, \& Pierre}]{Masoura2020}
Masoura, V.~A., Georgantopoulos, I., Mountrichas, G., {et~al.} 2020, Astronomy
  {\&} Astrophysics, 638, A45

\bibitem[{Masoura {et~al.}(2021)Masoura, Mountrichas, Georgantopoulos, \&
  Plionis}]{Masoura2021}
Masoura, V.~A., Mountrichas, G., Georgantopoulos, I., \& Plionis, M. 2021,
  Astronomy {\&} Astrophysics, 646, A167

\bibitem[{Masoura {et~al.}(2018)Masoura, Mountrichas, Georgantopoulos, Ruiz,
  Magdis, \& Plionis}]{Masoura2018}
Masoura, V.~A., Mountrichas, G., Georgantopoulos, I., {et~al.} 2018, A\&A, 618,
  31

\bibitem[{McCracken {et~al.}(2012)McCracken, Milvang-Jensen, Dunlop, Franx,
  Fynbo, F{\`{e}}vre, Holt, Caputi, Goranova, Buitrago, Emerson, Freudling,
  Hudelot, L{\'{o}}pez-Sanjuan, Magnard, Mellier, M{\o}ller, Nilsson,
  Sutherland, Tasca, \& Zabl}]{McCracken2012}
McCracken, H.~J., Milvang-Jensen, B., Dunlop, J., {et~al.} 2012, A\&A, 544,
  A156

\bibitem[{Menzel {et~al.}(2016)}]{Menzel2016}
Menzel, M.-L. {et~al.} 2016, MNRAS, 457, 110

\bibitem[{Merloni {et~al.}(2014)Merloni, Bongiorno, Brusa, Iwasawa, Mainieri,
  Magnelli, Salvato, Berta, Cappelluti, Comastri, Fiore, Gilli, \&
  Koekemoer}]{Merloni2014}
Merloni, A., Bongiorno, A., Brusa, M., {et~al.} 2014, Monthly Notices of the
  Royal Astronomical Society, 437, 3550

\bibitem[{Mountrichas \& Buat(2023)}]{Mountrichas2023d}
Mountrichas, G. \& Buat, V. 2023, Astronomy \& Astrophysics, 679, A151

\bibitem[{Mountrichas {et~al.}(2021{\natexlab{a}})Mountrichas, Buat,
  Georgantopoulos, Yang, Masoura, Boquien, \& Burgarella}]{Mountrichas2021b}
Mountrichas, G., Buat, V., Georgantopoulos, I., {et~al.} 2021{\natexlab{a}},
  Astronomy {\&} Astrophysics, 653, A70

\bibitem[{Mountrichas {et~al.}(2021{\natexlab{b}})Mountrichas, Buat, Yang,
  Boquien, Burgarella, \& Ciesla}]{Mountrichas2021a}
Mountrichas, G., Buat, V., Yang, G., {et~al.} 2021{\natexlab{b}}, Astronomy
  {\&} Astrophysics, 646, A29

\bibitem[{Mountrichas {et~al.}(2021{\natexlab{c}})Mountrichas, Buat, Yang,
  Boquien, Burgarella, Ciesla, Malek, \& Shirley}]{Mountrichas2021c}
Mountrichas, G., Buat, V., Yang, G., {et~al.} 2021{\natexlab{c}}, Astronomy
  {\&} Astrophysics, 653, A74

\bibitem[{Mountrichas {et~al.}(2022{\natexlab{a}})Mountrichas, Buat, Yang,
  Boquien, Burgarella, Ciesla, Malek, \& Shirley}]{Mountrichas2022b}
Mountrichas, G., Buat, V., Yang, G., {et~al.} 2022{\natexlab{a}}, Astronomy
  {\&} Astrophysics, 663, A130

\bibitem[{Mountrichas {et~al.}(2019)Mountrichas, Georgakakis, \&
  Georgantopoulos}]{Mountrichas2019}
Mountrichas, G., Georgakakis, A., \& Georgantopoulos, I. 2019, Monthly Notices
  of the Royal Astronomical Society, 483, 1374

\bibitem[{Mountrichas {et~al.}(2020)Mountrichas, Georgantopoulos, Ruiz, \&
  Kampylis}]{Mountrichas2020}
Mountrichas, G., Georgantopoulos, I., Ruiz, A., \& Kampylis, G. 2020, Monthly
  Notices of the Royal Astronomical Society, 491, 1727

\bibitem[{Mountrichas {et~al.}(2022{\natexlab{b}})Mountrichas, Masoura,
  Xilouris, Georgantopoulos, Buat, \& Paspaliaris}]{Mountrichas2022a}
Mountrichas, G., Masoura, V.~A., Xilouris, E.~M., {et~al.} 2022{\natexlab{b}},
  Astronomy {\&} Astrophysics, 661, A108

\bibitem[{Nenkova {et~al.}(2002)Nenkova, Ivezi{\'{c}}, \&
  Elitzur}]{Nenkova2002}
Nenkova, M., Ivezi{\'{c}}, {\v{Z}}., \& Elitzur, M. 2002, The Astrophysical
  Journal, 570, L9

\bibitem[{Netzer(2015)}]{Netzer2015}
Netzer, H. 2015, Annual Review of Astronomy and Astrophysics, 53, 365

\bibitem[{Ogawa {et~al.}(2021)Ogawa, Ueda, Tanimoto, \& Yamada}]{Ogawa2021}
Ogawa, S., Ueda, Y., Tanimoto, A., \& Yamada, S. 2021, The Astrophysical
  Journal, 906, 84

\bibitem[{Ordovas-Pascual {et~al.}(2017)Ordovas-Pascual, Mateos, Carrera,
  Wiersema, Barcons, Braito, Caccianiga, Del~Moro, Della~Ceca, \&
  Severgnini}]{OrdovasPascual2017}
Ordovas-Pascual, I., Mateos, S., Carrera, F.~J., {et~al.} 2017, Monthly Notices
  of the Royal Astronomical Society

\bibitem[{Park {et~al.}(2006)Park, Kashyap, Siemiginowska, van Dyk, Zezas,
  Heinke, \& Wargelin}]{Park2006}
Park, T., Kashyap, V.~L., Siemiginowska, A., {et~al.} 2006, The Astrophysical
  Journal, 652, 610

\bibitem[{{Pierre} {et~al.}(2016)}]{Pierre2016}
{Pierre}, M. {et~al.} 2016, A\&A, 592, 1

\bibitem[{Price-Whelan {et~al.}(2022)Price-Whelan, Lim, Earl, Starkman,
  Bradley, Shupe, Patil, Corrales, Brasseur, Nöthe, Donath, Tollerud, Morris,
  Ginsburg, Vaher, Weaver, Tocknell, Jamieson, van Kerkwijk, Robitaille, Merry,
  Bachetti, Günther, Aldcroft, Alvarado-Montes, Archibald, Bódi, Bapat,
  Barentsen, Bazán, Biswas, Boquien, Burke, Cara, Cara, Conroy, Conseil,
  Craig, Cross, Cruz, D'Eugenio, Dencheva, Devillepoix, Dietrich, Eigenbrot,
  Erben, Ferreira, Foreman-Mackey, Fox, Freij, Garg, Geda, Glattly,
  Gondhalekar, Gordon, Grant, Greenfield, Groener, Guest, Gurovich, Handberg,
  Hart, Hatfield-Dodds, Homeier, Hosseinzadeh, Jenness, Jones, Joseph,
  Kalmbach, Karamehmetoglu, Kałuszyński, Kelley, Kern, Kerzendorf, Koch,
  Kulumani, Lee, Ly, Ma, MacBride, Maljaars, Muna, Murphy, Norman, O'Steen,
  Oman, Pacifici, Pascual, Pascual-Granado, Patil, Perren, Pickering, Rastogi,
  Roulston, Ryan, Rykoff, Sabater, Sakurikar, Salgado, Sanghi, Saunders,
  Savchenko, Schwardt, Seifert-Eckert, Shih, Jain, Shukla, Sick, Simpson,
  Singanamalla, Singer, Singhal, Sinha, Sipőcz, Spitler, Stansby, Streicher,
  Šumak, Swinbank, Taranu, Tewary, Tremblay, de~Val-Borro, Kooten, Vasović,
  Verma, de~Miranda~Cardoso, Williams, Wilson, Winkel, Wood-Vasey, Xue,
  Yoachim, ZHANG, \& Zonca}]{PriceWhelan2022}
Price-Whelan, A.~M., Lim, P.~L., Earl, N., {et~al.} 2022, ApJ
  [\eprint[arXiv]{2206.14220}]

\bibitem[{Ricci {et~al.}(2017)Ricci, Trakhtenbrot, Koss, Ueda, Vecchio,
  Treister, Schawinski, Paltani, Oh, Lamperti, Berney, Gandhi, Ichikawa, Bauer,
  Ho, Asmus, Beckmann, Soldi, Balokovi{\'{c}}, Gehrels, \&
  Markwardt}]{Ricci2017}
Ricci, C., Trakhtenbrot, B., Koss, M.~J., {et~al.} 2017, The Astrophysical
  Journal Supplement Series, 233, 17

\bibitem[{Ruiz {et~al.}(2018)Ruiz, Corral, Mountrichas, \&
  Georgantopoulos}]{Ruiz2018}
Ruiz, A., Corral, A., Mountrichas, G., \& Georgantopoulos, I. 2018, Astronomy
  {\&} Astrophysics, 618, A52

\bibitem[{Salvato {et~al.}(2018)Salvato, Buchner, Budav{\'{a}}ri, Dwelly,
  Merloni, Brusa, Rau, Fotopoulou, \& Nandra}]{Salvato2018b}
Salvato, M., Buchner, J., Budav{\'{a}}ri, T., {et~al.} 2018, Monthly Notices of
  the Royal Astronomical Society, 473, 4937

\bibitem[{Salvato {et~al.}(2009)Salvato, Hasinger, Ilbert, Zamorani, Brusa,
  Scoville, Rau, Capak, Arnouts, Aussel, Bolzonella, Buongiorno, Cappelluti,
  Caputi, Civano, Cook, Elvis, Gilli, Jahnke, Kartaltepe, Impey, Lamareille,
  Le~Floc'h, Lilly, Mainieri, McCarthy, McCracken, Mignoli, Mobasher, Murayama,
  Sasaki, S~anders, Schiminovich, Shioya, Shopbell, Silverman, Smol{\v
  c}i{\'c}, Surace, Taniguchi, Thompson, Trump, Urry, \&
  Zamojski}]{Salvato2009}
Salvato, M., Hasinger, G., Ilbert, O., {et~al.} 2009, ApJ, 690, 1250

\bibitem[{Salvato {et~al.}(2022)Salvato, Wolf, Dwelly, Georgakakis, Brusa,
  Merloni, Liu, Toba, Nandra, Lamer, Buchner, Schneider, Freund, Rau, Schwope,
  Nishizawa, Klein, Arcodia, Comparat, Musiimenta, Nagao, Brunner, Malyali,
  Finoguenov, Anderson, Shen, Ibarra-Medel, Trump, Brandt, Urry, Rivera,
  Krumpe, Urrutia, Miyaji, Ichikawa, Schneider, Fresco, Boller, Haase,
  Brownstein, Lane, Bizyaev, \& Nitschelm}]{Salvato2022}
Salvato, M., Wolf, J., Dwelly, T., {et~al.} 2022, Astronomy \& Astrophysics,
  661, A3

\bibitem[{{Salvato} {et~al.}(2011)}]{Salvato2011}
{Salvato}, M. {et~al.} 2011, ApJ, 742, 61

\bibitem[{{Scoville} {et~al.}(2007)}]{Scoville2007}
{Scoville}, N. {et~al.} 2007, ApJS, 172, 1

\bibitem[{Shimizu {et~al.}(2018)Shimizu, Davies, Koss, Ricci, Lamperti, Oh,
  Schawinski, Trakhtenbrot, Burtscher, Genzel, Lin, Lutz, Rosario, Sturm, \&
  Tacconi}]{Shimizu2018}
Shimizu, T.~T., Davies, R.~I., Koss, M., {et~al.} 2018, The Astrophysical
  Journal, 856, 154

\bibitem[{Shirley {et~al.}(2021)Shirley, Duncan, Varillas, Hurley, Ma{\l}ek,
  Roehlly, Smith, Aussel, Bakx, Buat, Burgarella, Christopher, Duivenvoorden,
  Eales, Efstathiou, Solares, Griffin, Jarvis, Faro, Marchetti, McCheyne,
  Papadopoulos, Penner, Pons, Prescott, Rigby, Rottgering, Saxena, Scudder,
  Vaccari, Wang, \& Oliver}]{Shirley2021}
Shirley, R., Duncan, K., Varillas, M. C.~C., {et~al.} 2021, MNRAS, 507, 129

\bibitem[{Shirley {et~al.}(2019)Shirley, Roehlly, Hurley, Buat, del Carmen
  Campos~Varillas, Duivenvoorden, Duncan, Efstathiou, Farrah, Solares, Malek,
  Marchetti, McCheyne, Papadopoulos, Pons, Scipioni, Vaccari, \&
  Oliver}]{Shirley2019}
Shirley, R., Roehlly, Y., Hurley, P.~D., {et~al.} 2019, Monthly Notices of the
  Royal Astronomical Society, 490, 634

\bibitem[{Stalevski {et~al.}(2012)Stalevski, Fritz, Baes, Nakos, \&
  Popovi{\'{c}}}]{Stalevski2012}
Stalevski, M., Fritz, J., Baes, M., Nakos, T., \& Popovi{\'{c}}, L.~{\v{C}}.
  2012, Monthly Notices of the Royal Astronomical Society, 420, 2756

\bibitem[{Stalevski {et~al.}(2016)Stalevski, Ricci, Ueda, Lira, Fritz, \&
  Baes}]{Stalevski2016}
Stalevski, M., Ricci, C., Ueda, Y., {et~al.} 2016, Monthly Notices of the Royal
  Astronomical Society, 458, 2288

\bibitem[{{Sutherland} \& {Saunders}(1992)}]{Sutherland_and_Saunders1992}
{Sutherland}, W. \& {Saunders}, W. 1992, MNRAS, 259, 413

\bibitem[{Taylor(2005)}]{Taylor2005}
Taylor, M.~B. 2005, in Astronomical Society of the Pacific Conference Series,
  Vol. 347, Astronomical Data Analysis Software and Systems XIV, ed.
  P.~{Shopbell}, M.~{Britton}, \& R.~{Ebert}, 29

\bibitem[{Trippe {et~al.}(2010)Trippe, Crenshaw, Deo, Dietrich, Kraemer,
  Rafter, \& Turner}]{Trippe2010}
Trippe, M.~L., Crenshaw, D.~M., Deo, R.~P., {et~al.} 2010, The Astrophysical
  Journal, 725, 1749

\bibitem[{Urry \& Padovani(1995)}]{Urry1995}
Urry, C.~M. \& Padovani, P. 1995, Publications of the Astronomical Society of
  the Pacific, 107, 803

\bibitem[{Villa-Velez {et~al.}(2021)Villa-Velez, Buat, Theule, Boquien, \&
  Burgarella}]{VillaVelez2021}
Villa-Velez, J.~A., Buat, V., Theule, P., Boquien, M., \& Burgarella, D. 2021,
  Astronomy \& Astrophysics, 654, A153

\bibitem[{Whittle(1992)}]{Whittle1992}
Whittle, M. 1992, The Astrophysical Journal Supplement Series, 79, 49

\bibitem[{{Wright} {et~al.}(2010){Wright}, {Eisenhardt}, {Mainzer}, {Ressler},
  {Cutri}, {Jarrett}, {Kirkpatrick}, {Padgett}, {McMillan}, {Skrutskie},
  {Stanford}, {Cohen}, {Walker}, {Mather}, {Leisawitz}, {Gautier}, {McLean},
  {Benford}, {Lonsdale}, {Blain}, {Mendez}, {Irace}, {Duval}, {Liu}, {Royer},
  {Heinrichsen}, {Howard}, {Shannon}, {Kendall}, {Walsh}, {Larsen}, {Cardon},
  {Schick}, {Schwalm}, {Abid}, {Fabinsky}, {Naes}, \& {Tsai}}]{Wright2010}
{Wright}, E.~L., {Eisenhardt}, P.~R.~M., {Mainzer}, A.~K., {et~al.} 2010, AJ,
  140, 1868

\bibitem[{Yang {et~al.}(2022)Yang, Boquien, Brandt, Buat, Burgarella, Ciesla,
  Lehmer, Małek, Mountrichas, Papovich, Pons, Stalevski, Theulé, \&
  Zhu}]{Yang2022}
Yang, G., Boquien, M., Brandt, W.~N., {et~al.} 2022, A\&A [\eprint{2201.03718}]

\bibitem[{Yang {et~al.}(2020)Yang, Boquien, Buat, Burgarella, Ciesla, Duras,
  Stalevski, Brandt, \& Papovich}]{Yang2020}
Yang, G., Boquien, M., Buat, V., {et~al.} 2020, Monthly Notices of the Royal
  Astronomical Society, 491, 740

\bibitem[{{Yang} {et~al.}(2017){Yang}, {Chen}, {Vito}, {Brandt}, {Alexander},
  {Luo}, {Sun}, {Xue}, {Bauer}, {Koekemoer}, {Lehmer}, {Liu}, {Schneider},
  {Shemmer}, {Trump}, {Vignali}, \& {Wang}}]{Yang2017}
{Yang}, G., {Chen}, C. T.~J., {Vito}, F., {et~al.} 2017, ApJ, 842, 72

\bibitem[{Zou {et~al.}(2019)Zou, Yang, Brandt, \& Xue}]{Zou2019}
Zou, F., Yang, G., Brandt, W.~N., \& Xue, Y. 2019, The Astrophysical Journal,
  878, 11

\end{thebibliography}
\bibliographystyle{aa}

\end{document}